\documentclass[aps, pra, twocolumn, preprintnumbers, amsmath, amssymb, superscriptaddress, notitlepage]{revtex4}

\usepackage{url}
\usepackage{braket}
\usepackage{graphicx}
\usepackage{multirow}
\usepackage{amsthm}
\usepackage{soul}
\usepackage{verbatim}
\usepackage{bm}
\usepackage{outlines}
\usepackage[exponent-product=\cdot]{siunitx}
\usepackage[space]{grffile}
\usepackage[dvipsnames]{xcolor}
\usepackage{xr}
\usepackage{color}
\usepackage{tikz}
\usetikzlibrary{calc, positioning}
\usepackage{mathtools}
\usepackage{algorithm}
\usepackage{algorithmic}
\usepackage{amsmath}
\usepackage{amssymb}
\usepackage{comment}
\usepackage[english]{babel}
\usepackage{subfigure}
\usepackage{booktabs}

\usepackage{xifthen}

\newcommand\myshade{85}
\colorlet{mylinkcolor}{violet}
\colorlet{mycitecolor}{YellowOrange}
\colorlet{myurlcolor}{Aquamarine}
\usepackage{hyperref}
\hypersetup{
	linkcolor  = mylinkcolor!\myshade!black,
	citecolor  = mycitecolor!\myshade!black,
	urlcolor   = myurlcolor!\myshade!black,
	colorlinks = true,
}
\usepackage[capitalize]{cleveref}

\newcommand{\MHz}{\mathrm{MHz}}

\newcommand{\ns}{\mathrm{ns}}
\newcommand{\us}{\mu\mathrm{s}}

\newcommand{\degrees}{\mathrm{deg}}

\newcommand{\mean}[1]{\overline{#1}}
\newcommand{\quotes}[1]{``#1''}
\newcommand{\abs}[1]{\left|{#1}\right|}
\newcommand{\brkt}[1]{\left(#1\right)}
\newcommand{\bbrkt}[1]{\left[#1\right]}
\newcommand{\com}[1]{\left[#1\right]}
\newcommand{\setbrkt}[1]{\left\{#1\right\}}
\newcommand{\anticom}[1]{\left\{#1\right\}}

\newcommand{\imnum}{i}
\newcommand{\dmsymbol}{\rho}
\newcommand{\hamil}[1][]{\ifthenelse{\equal{#1}{}}{H}{H_{\mathrm{#1}}}}
\newcommand{\lindbladop}[1][]{\ifthenelse{\equal{#1}{}}{L}{L_{\mathrm{#1}}}}

\newcommand{\logerrrate}{\varepsilon_{\mathrm{L}}}
\newcommand{\logfid}[1][]{\ifthenelse{\equal{#1}{}}{F_{\mathrm{L}}}{F_{\mathrm{L}}\left(#1\right)}}
\newcommand{\fid}[1][]{\ifthenelse{\equal{#1}{}}{F}{F_{\mathrm{#1}}}}
\newcommand{\errrate}[1][]{\ifthenelse{\equal{#1}{}}{\varepsilon}{\varepsilon_{\mathrm{#1}}}}

\newcommand{\qubit}[1][]{\ifthenelse{\equal{#1}{}}{Q}{Q_{\mathrm{#1}}}}
\newcommand{\qubitind}[1][]{\ifthenelse{\equal{#1}{}}{Q}{Q_{#1}}}

\newcommand{\leakrate}{L_{1}}
\newcommand{\seeprate}{L_{2}}

\newcommand{\tcycle}{t_{\text{c}}}
\newcommand{\tcz}{t_{\CZ}}

\newcommand{\Tone}{T_{1}}
\newcommand{\tdephsweet}{T_{\phi, \text{max}}}
\newcommand{\tdephint}{T_{\phi, \text{int}}}
\newcommand{\tdephpark}{T_{\phi, \text{park}}}

\newcommand{\tint}{t_{\mathrm{int}}}
\newcommand{\tphasecorr}{t_{\mathrm{cor}}}

\newcommand{\lcphase}[1][]{\ifthenelse{\equal{#1}{}}{\phi^{\mathrm{L}}}{\phi^{\mathrm{L}}_{\mathrm{#1}}}}

\newcommand{\identity}{I}
\newcommand{\leakyI}{\tilde{I}}
\newcommand{\leakyZ}{\tilde{Z}}
\newcommand{\leakyX}{\tilde{X}}

\newcommand{\coupling}{J_1}

\newcommand{\cycle}[1][]{\ifthenelse{\equal{#1}{}}{n}{n_{\mathrm{#1}}}}
\newcommand{\signset}[1][]{\ifthenelse{\equal{#1}{}}{\mathcal{S}}{\mathcal{S}_{\mathrm{#1}}}}
\newcommand{\ztype}{Z}
\newcommand{\xtype}{X}
\newcommand{\datatype}{D}
\newcommand{\anctype}{A}

\newcommand{\dataq}[1]{D_{\text{#1}}}

\newcommand{\zanc}[1]{Z_{\text{#1}}}

\newcommand{\probdefect}[1][]{\ifthenelse{\equal{#1}{}}{p^{d}}{p^{d}_{#1}}}

\newcommand{\proberrleaked}[1][]{\ifthenelse{\equal{#1}{}}{p_{\textrm{d}}^{l}}{p_{\textrm{d}}^{l}\left(\mathrm{#1}\right)}}
\newcommand{\meanproberrleaked}[1][]{\ifthenelse{\equal{#1}{}}{\mean{p}_{\textrm{d}}^{l}}{\mean{p}_{\textrm{d}}^{l}\left(\mathrm{#1}\right)}}
\newcommand{\numsignatures}[1][]{\ifthenelse{\equal{#1}{}}{N_{\mathcal{S}}}{N_{\mathcal{S}\mathrm{,#1}}}}
\newcommand{\tgate}{t_{\mathrm{single}}}

\newcommand{\HMM}{\text{HMM}}
\newcommand{\HMMs}{HMMs}
\newcommand{\DM}{\text{DM}}

\newcommand{\threshold}{\mathrm{th}}

\newcommand{\probleak}[1][]{\ifthenelse{\equal{#1}{}}{p^{\mathcal{L}}}{p^{\mathcal{L}}_{#1}}}
\newcommand{\probcomp}[1][]{\ifthenelse{\equal{#1}{}}{p^{\mathcal{C}}}{p^{\mathcal{C}}_{#1}}}

\newcommand{\loggrowth}[1][]{\ifthenelse{\equal{#1}{}}{g}{g_{\mathrm{#1}}}}
\newcommand{\precision}[1][]{\ifthenelse{\equal{#1}{}}{\mathcal{P}}{\mathcal{P}_{\mathrm{#1}}}}
\newcommand{\recall}[1][]{\ifthenelse{\equal{#1}{}}{\mathcal{R}}{\mathcal{R}_{\mathrm{#1}}}}
\newcommand{\prob}[1][]{\ifthenelse{\equal{#1}{}}{\mathbb{P}}{\mathbb{P}_{#1}}}
\newcommand{\auc}[1][]{\ifthenelse{\equal{#1}{}}{\mathrm{AUC}}{\mathrm{AUC}_{#1}}}
\newcommand{\hmmopt}[1][]{\ifthenelse{\equal{#1}{}}{\mathcal{O}}{\mathcal{O}_{\mathrm{#1}}}}
\newcommand{\leakassfid}[1][]{\ifthenelse{\equal{#1}{}}{e^{\mathcal{L}}}{e^{\mathcal{L}}_{\mathrm{#1}}}}
\newcommand{\leakmeasfid}[1][]{\ifthenelse{\equal{#1}{}}{F^{\mathcal{L}}}{F^{\mathcal{L}}_{#1}}}

\newcommand{\dataind}{D}

\newcommand{\gaussian}[1][]{\ifthenelse{\equal{#1}{}}{\mathcal{N}}{\mathcal{N}_{#1}}}
\newcommand{\gausmean}[1][]{\ifthenelse{\equal{#1}{}}{\mu}{\mu_{#1}}}
\newcommand{\gausstd}[1][]{\ifthenelse{\equal{#1}{}}{\sigma}{\sigma_{#1}}}
\newcommand{\pdf}[1][]{\ifthenelse{\equal{#1}{}}{f}{f_{\mathrm{#1}}}}
\newcommand{\meas}[1][]{\ifthenelse{\equal{#1}{}}{m}{m\left[#1\right]}}

\newcommand{\inphasecomp}[1][]{\ifthenelse{\equal{#1}{}}{I}{I_{#1}}}
\newcommand{\quadraturecomp}[1][]{\ifthenelse{\equal{#1}{}}{Q}{Q_{#1}}}

\newcommand{\surfaceprobleaks}[1][]{\ifthenelse{\equal{#1}{}}{\vec{p}^{\mathcal{L}}}{\vec{p}^{\mathcal{L}}_{\mathrm{#1}}}}
\newcommand{\surfaceprobcomps}[1][]{\ifthenelse{\equal{#1}{}}{\vec{p}^{\mathcal{C}}}{\vec{p}^{\mathcal{C}}_{\mathrm{#1}}}}
\newcommand{\datafrac}[1][]{\ifthenelse{\equal{#1}{}}{f}{f_{\mathrm{#1}}}}

\newcommand{\tmeas}{t_{\mathrm{\meas}}}

\newcommand{\flux}{\mathrm{flux}}
\newcommand{\static}{\mathrm{stat}}
\newcommand{\sweet}{\mathrm{max}}
\newcommand{\interact}{\mathrm{int}}
\newcommand{\anharm}{\alpha}

\newcommand{\CZ}{\mathrm{CZ}}

\newcommand{\leakcondphase}[1][]{\ifthenelse{\equal{#1}{}}{\phi^{\leaksub}}{\phi^{\leaksub}_{\mathrm{#1}}}}

\newcommand{\compsub}{\mathcal{C}}
\newcommand{\leaksub}{\mathcal{L}}
\newcommand{\leakmobility}{L_{\mathrm{m}}}
\newcommand{\superleak}{L_3}

\newcommand{\Tdeph}{T_{\phi}}

\newcommand{\logstate}[1]{\ket{#1}_{\mathrm{L}}}
\newcommand{\leaktime}[1][]{\ifthenelse{\equal{#1}{}}{t^{\mathcal{L}}}{t^{\mathcal{L}}_{\mathrm{#1}}}}

\newcommand{\be}{\begin{equation}}
\newcommand{\ee}{\end{equation}}
\newcommand{\ba}{\begin{array}}
	\newcommand{\ea}{\end{array}}
\newcommand{\bea}{\begin{eqnarray}}
\newcommand{\eea}{\end{eqnarray}}

\newcommand{\QEC}{QEC}
\newcommand{\qecround}[1][]{\ifthenelse{\equal{#1}{}}{\vec{t}}{t_{\mathrm{#1}}}}
\newcommand{\measvec}[1][]{\ifthenelse{\equal{#1}{}}{\vec{m}}{\vec{m}_{\mathrm{#1}}}}
\newcommand{\stabvec}[1][]{\ifthenelse{\equal{#1}{}}{\vec{s}}{\vec{s}_{\mathrm{#1}}}}
\newcommand{\synvec}[1][]{\ifthenelse{\equal{#1}{}}{\vec{s}}{\vec{s}_{\mathrm{#1}}}}
\newcommand{\defvec}[1][]{\ifthenelse{\equal{#1}{}}{\vec{d}}{\vec{d}_{\mathrm{#1}}}}
\newcommand{\defect}[1][]{\ifthenelse{\equal{#1}{}}{d}{d\left[#1\right]}}
\newcommand{\syndrome}[1][]{\ifthenelse{\equal{#1}{}}{s}{s\left[#1\right]}}
\newcommand{\errorvec}[1][]{\ifthenelse{\equal{#1}{}}{\vec{e}}{\vec{e}_{\mathrm{#1}}}}
\newcommand{\hiddenvec}[1][]{\ifthenelse{\equal{#1}{}}{\vec{l}}{\vec{l}_{\mathrm{#1}}}}

\newcommand{\SNR}{\mathrm{SNR}}
\newcommand{\erfc}{\mathrm{erfc}}

\newcommand{\logical}{\mathrm{L}}
\newcommand{\freq}{\omega}

\newcommand{\ptmop}{R}
\newcommand{\ampdamp}{\downarrow}
\newcommand{\gate}{\text{gate}}
\newcommand{\idle}{\text{idle}}

\newcommand{\state}{s}
\newcommand{\obs}{o}
\newcommand{\obsi}{o_{i}}
\newcommand{\probstate}[1][]{\ifthenelse{\equal{#1}{}}{p^{\state}}{p^{\state}_{#1}}}
\newcommand{\probstateprime}[1][]{\ifthenelse{\equal{#1}{}}{p^{\state'}}{p^{\state'}_{#1}}}
\newcommand{\probobs}[1][]{\ifthenelse{\equal{#1}{}}{p^{\obs}}{p^{\obs}_{#1}}}

\begin{document}

	\title{Leakage detection for a transmon-based surface code}

	\newcommand{\QuTech}{\affiliation{QuTech, Delft University of Technology, P.O.~Box 5046, 2600 GA Delft, The Netherlands}}
	\newcommand{\Kavli}{\affiliation{Kavli Institute of Nanoscience, Delft University of Technology, P.O.~Box 5046, 2600 GA Delft, The Netherlands}}
	\newcommand{\JARA}{\affiliation{JARA Institute for Quantum Information, Forschungszentrum Juelich, D-52425 Juelich, Germany}}
	\newcommand{\TNO}{\affiliation{Netherlands Organisation for Applied Scientiﬁc Research (TNO), P.O.~Box 96864, 2509 JG The Hague, The Netherlands}}
	\newcommand{\lorentz}{\affiliation{Instituut-Lorentz, Universiteit Leiden, P.O.~Box 9506, 2300 RA Leiden, The Netherlands}}
	\newcommand{\google}{\affiliation{Google Research, Venice, CA 90291, United States}}

	\author{B.~M.~Varbanov}\QuTech
	\author{F.~Battistel}\QuTech
	\author{B.~M.~Tarasinski}\QuTech\Kavli
	\author{V.~P.~Ostroukh}\QuTech\Kavli
	\author{T.~E.~O'Brien}\lorentz\google
	\author{L.~DiCarlo}\QuTech\Kavli
	\author{B.~M.~Terhal}\QuTech\JARA

	\date{\today}

	\begin{abstract}
		Leakage outside of the qubit computational subspace, present in many leading experimental platforms, constitutes a threatening error for quantum error correction~(QEC) for qubits.
		We develop a leakage-detection scheme via Hidden Markov models~(HMMs) for transmon-based implementations of the surface code.
		By performing realistic density-matrix simulations of the distance-3 surface code (Surface-17), we observe that leakage is sharply projected and leads to an increase in the surface-code defect probability of neighboring stabilizers.
		Together with the analog readout of the ancilla qubits, this increase enables the accurate detection of the time
		and location of leakage.
		We restore the logical error rate below the memory break-even point by post-selecting out leakage, discarding about~\(47\%\) of the data.
		Leakage detection via HMMs opens the prospect for near-term QEC demonstrations, targeted leakage reduction and leakage-aware decoding and is applicable to other experimental platforms.
	\end{abstract}

	\maketitle

	Recent advances in qubit numbers~\cite{Corcoles19,Arute19,Otterbach17,Landsman19}, as well as operational~\cite{Barends14,Barends19,Rol16,Chen16b,Rol19a,Foxen20,Hong19,Sheldon16b,Harty14} and measurement~\cite{Jeffrey14,Bultink16,Heinsoo18} fidelities have enabled leading quantum computing platforms, such as superconducting and trapped-ion processors, to target demonstrations of quantum error correction~(QEC)~\cite{Kelly15,Riste15,Takita16,Bultink19,Andersen19,Negnevitsky18,Andersen19b} and quantum advantage~\cite{Boixo18,Neill18,Arute19,Bravyi18}.
	In particular, two-dimensional stabilizer codes, such as the surface code, are a promising approach~\cite{Obrien17,Andersen19b} towards achieving quantum fault tolerance and, ultimately, large-scale quantum computation~\cite{Fowler12}.
	One of the central assumptions of textbook~QEC is that any error can be decomposed into a set of Pauli errors that act within the computational space of the qubit.
	In practice, many qubits such as weakly-anharmonic transmons, as well as hyperfine-level trapped ions, are many-level systems which function as qubits by restricting the interactions with the other excited states.
	Due to imprecise control~\cite{Chow11,Sheldon16b,Tripathi19} or the explicit use of non-computational states for operations~\cite{Strauch03,DiCarlo09,Martinis14,Rol19a,Barends14,Barends19,Caldwell18,Hong19,Ghosh13}, there exists a finite probability for information to leak from the computational subspace.
	Thus, leakage constitutes an error that falls outside of the domain of the qubit stabilizer formalism.
	Furthermore, leakage can last over many QEC~cycles, despite having a finite duration set by the relaxation time~\cite{Ghosh13_B}.
	Hence, leakage represents a menacing error source in the context of quantum error correction~\cite{Aliferis07,Fowler13,Ghosh13_B,Ghosh15,Kelly15,Suchara15,Brown18,Brown19,Brown19_B}, despite leakage probabilities per operation being smaller than readout, control or decoherence error probabilities~\cite{Rol19a,Barends19,Chen16b,Motzoi09}.

	The presence of leakage errors has motivated investigations of its effect on the code performance and of strategies to mitigate it.
	A number of previous studies have focused on a stochastic depolarizing model of leakage~\cite{Fowler13,Suchara15,Brown18,Brown19,Brown19_B}, allowing to explore large-distance surface codes and the reduction of the code threshold using simulations.
	These models, however, do not capture the full details of leakage, even though a specific adaptation has been used in the case of trapped-ion qubits~\cite{Brown18,Brown19,Brown19_B}.
	Complementary studies have considered a physically realistic leakage model for transmons~\cite{Ghosh13_B,Ghosh15}, which was only applied to a small parity-check unit due to the computational cost of many-qutrit density-matrix simulations.
	In either case, leakage was found to have a strong impact on the performance of the code, resulting in the propagation of errors, in the increase of the logical error rate and in a reduction of the effective code distance.
	In order to mitigate these effects, there have been proposals for the introduction of leakage reduction units~(LRUs)~\cite{Aliferis07,Suchara15,Ghosh15,Hayes19} beyond the natural relaxation channel, for modifications to the decoding algorithms~\cite{Fowler13,Suchara15,Kelly15}, as well as for the use of different codes altogether~\cite{Brown19}.
	Many of these approaches rely on the detection of leakage or introduce an overhead in the execution of the code.
	Recently, the indirect detection of leakage in a 3-qubit parity-check experiment~\cite{Bultink19} was realized via a Hidden Markov Model~(HMM), allowing for subsequent mitigation via post-selection.
	Given that current experimental platforms are within reach of quantum-memory demonstrations, detailed simulations employing realistic leakage models are vital for a comprehensive understanding of the effect of leakage on the code performance, as well as for the development of a strategy to detect leakage without additional overhead.

	In this work we demonstrate the use of computationally efficient~HMMs to detect leakage in a transmon implementation of the distance-3 surface code (Surface-17)~\cite{Versluis17}.
	Using full-density-matrix simulations~\cite{Obrien17,quantumsim_website} we first show that repeated stabilizer measurements sharply project data qubits into the leakage subspace, justifying the use of classical~HMMs with only two hidden states (computational or leaked) for leakage detection.
	We observe a considerable increase in the surface-code defect probability of neighboring stabilizers while a data or ancilla qubit is leaked, a clear signal that may be detected by the~HMMs.
	For ancilla qubits, we further consider the information available in the analog measurement outcomes, even when the leaked state~\(\ket{2}\) can be discriminated from the computational states~\(\ket{0}\) and~\(\ket{1}\) with limited fidelity.
	We demonstrate that a set of two-state~HMMs, one~HMM for each qubit, can accurately detect both the time and the location of a leakage event in the surface code.
	By post-selecting on the detected leakage, we restore the logical performance of Surface-17 (using targeted experimental parameters) below the memory break-even point, while discarding only~\(\approx47\%\) of the data.
	Finally, we outline a minimal set of conditions for our leakage-detection scheme to apply to other quantum-computing platforms.
	These results open the prospect for near-term demonstrations of fault tolerance even in the presence of leakage, as well as leakage-aware decoding~\cite{Kelly15,Suchara15} and real-time targeted application of LRU~schemes~\cite{Aliferis07,Suchara15,Ghosh15}.

	\section{Results}
	\label{sec:results}

	\subsection{Leakage error model}
	\label{sub:leakage_error_model}

	\begin{figure}
		\centering
		\includegraphics[width=\columnwidth]{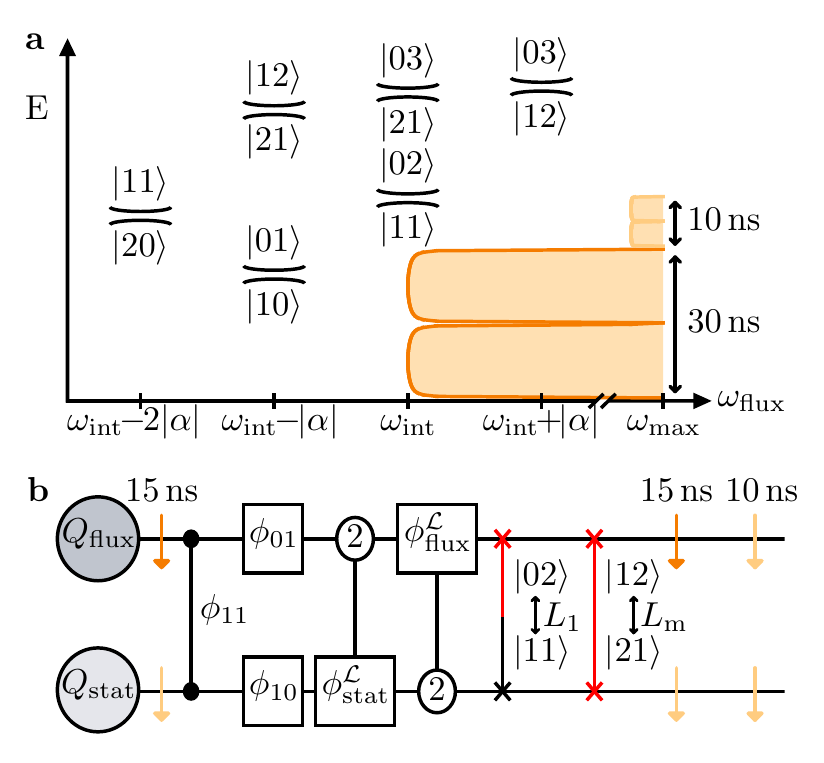}
		\caption{\label{fig:figure1_CZmodel}
			Schematic of the relevant interactions and the $\CZ$~error model for two transmons.
			\textbf{a}~Sketch of the considered avoided crossings, arranged vertically versus energy and horizontally versus the frequency~$\freq_\flux$ of~$\qubit[\flux]$.
			Inset: Schematic diagram of the frequency excursion taken by~$\qubit[\flux]$ during a Net-Zero pulse~\cite{Rol19a}.
			\textbf{b}~The parametrized $\CZ$ error model.
			Phase errors, SWAP-like exchanges, relaxation and decoherence are taken into account~(see~\cref{sub:leakage_error_model}).
		}
	\end{figure}

	We develop an error model for leakage in superconducting transmons, for which two-qubit gates constitute the dominant source of leakage~\cite{Strauch03,DiCarlo09,Martinis14,Rol19a,Barends14,Barends19,Caldwell18,Hong19,Chow11,Sheldon16b,Tripathi19}, while single-qubit gates have negligible leakage probabilities~\cite{Chen16b,Motzoi09}.
	We thus focus on the former, while the latter is assumed to induce no leakage at all.
	We assume that single-qubit gates act on a leaked state as the identity.
	Measurement-induced leakage is also assumed to be negligible.

	We use full-trajectory simulations to characterize leakage in the Net-Zero implementation~\cite{Rol19a} of the controlled-phase gate~($\CZ$), considered as the native two-qubit gate in a transmon-based Surface-17, with experimentally targeted parameters (see~\cref{tab:sim_params} and~\cref{tab:target_parameters}).
	This gate uses a flux pulse such that the higher frequency qubit~($\qubit[\flux]$) is fluxed down from its sweetspot frequency~$\freq_{\sweet}$ to the vicinity of the interaction frequency~$\freq_{\interact}=\freq_{\static}-\anharm$, where $\freq_{\static}$~is the frequency of the other qubit~($\qubit[\static]$), which remains static, and $\alpha$ is the transmon anharmonicity.
	The inset in~\cref{fig:figure1_CZmodel}~\textbf{a} shows a schematic diagram of the frequency excursion taken by~$\qubit[\flux]$.
	A (bipolar) $30~\ns$~pulse tunes twice the qubit on resonance with the $\ket{11}\leftrightarrow\ket{02}$~avoided crossing, corresponding to the interaction frequency~$\freq_{\interact}$.
	This pulse is followed by a pair of $10~\ns$~single-qubit phase-correction pulses.
	The relevant crossings around~$\freq_{\interact}$ are shown in~\cref{fig:figure1_CZmodel}~\textbf{a} and are all taken into account in the full-trajectory simulations. The two-qubit interactions give rise to population exchanges towards and within the leakage subspace and to the phases acquired during gates with leaked qubits, which we model as follows.

	The model in~\cref{fig:figure1_CZmodel}~\textbf{b} considers a general $\CZ$~rotation, characterized by the two-qubit phase~$\phi_{11}$ for state~$\ket{11}$ and~$\phi=0$ for the other three computational states.
	The single-qubit relative phases~$\phi_{01}$ and~$\phi_{10}$ result from imperfections in the phase corrections.
	The conditional phase is defined as~$\phi_{\CZ}=\phi_{11}-\phi_{01}-\phi_{10}+\phi_{00}$, which for an ideal~\(\CZ\) is~$\phi_{\CZ}=\pi$.
	In this work, we assume~\(\phi_{00}=\phi_{01}=\phi_{10}=0\) and~\(\phi_{\CZ}=\phi_{11}=\pi\).
	We set~$\phi_{02}=-\phi_{11}$ in the rotating frame of the qutrit, as it holds for flux-based gates~\cite{Ghosh13}.

	Interactions between leaked and non-leaked qubits lead to extra phases, which we call leakage conditional phases.
	We consider first the interaction between a leaked~\(\qubit[\flux]\) and a non-leaked~\(\qubit[\static]\).
	In this case the gate restricted to the \(\left\{\ket{02},\ket{12}\right\}\)~subspace has the effect ${diag}\brkt{e^{i\phi_{02}},e^{i\phi_{12}}}$,
	which up to a global phase corresponds to a $Z$~rotation on~\(\qubit[\static]\) with an angle given by the leakage conditional phase $\leakcondphase[stat]\coloneqq \phi_{02}-\phi_{12}$.
	Similarly, if~\(\qubit[\static]\) is leaked, then~\(\qubit[\flux]\) acquires a leakage conditional phase $\leakcondphase[flux]\coloneqq \phi_{20}-\phi_{21}$.
	These rotations are generally non-trivial, i.e.,~$\leakcondphase[stat]\neq \pi$ and~$\leakcondphase[flux]\neq 0$, due to the interactions in the 3-excitation manifold which cause a shift in the energy of~$\ket{12}$ and~$\ket{21}$ (see~\cref{sec:dynamics_leakage_subspace}).
	If the only interaction leading to non-trivial~$\leakcondphase[stat]$,~$\leakcondphase[flux]$ is the interaction between~\(\ket{12}\) and~\(\ket{21}\), then it can be expected that~$\phi_{12}=-\phi_{21}$ in the rotating frame of the qutrit, leading to $\leakcondphase[stat]=\pi - \leakcondphase[flux]$.

	Leakage is modeled as an exchange between~$\ket{11}$ and~$\ket{02}$, i.e.,~\(\ket{11}\mapsto \sqrt{1-4\leakrate}\ket{11} + e^{i\phi}\sqrt{4\leakrate}\ket{02}\) and \(\ket{02}\mapsto -e^{-i\phi}\sqrt{4\leakrate}\ket{11} + \sqrt{1-4\leakrate}\ket{02}\), with~\(\leakrate\) the leakage probability~\cite{Wood18}.
	We observe that the phase~\(\phi\) and the off-diagonal elements~\(\ket{11}\bra{02}\) and~\(\ket{02}\bra{11}\) do not affect the results presented in this work, so we set them to~0 for computational efficiency (see~\cref{sec:sim_parameters}).
	The SWAP-like exchange between~\(\ket{12}\) and~\(\ket{21}\) with probability~\(\leakmobility\), which we call leakage mobility, as well as the possibility of further leaking to~$\ket{3}$, are analyzed in~\cref{sec:dynamics_leakage_subspace}.

	The described operations are implemented in \emph{quantumsim}~\cite{quantumsim_website} as instantaneous, while the amplitude and phase damping experienced by the transmon during the application of the gate are symmetrically introduced around them, indicated by light-orange arrows in~\cref{fig:figure1_CZmodel}~\textbf{b}.
	The dark-orange arrows indicate the increased dephasing rate of~$\qubit[\flux]$ far away from~\(\freq_{\sweet}\) during the Net-Zero pulse.
	The error parameters considered in this work are summarized in~\cref{sec:sim_parameters}.
	In particular, unless otherwise stated, $\leakrate$~is set to~$0.125\%$ and $\leakcondphase[flux]$ and~$\leakcondphase[stat]$ are randomized for each qubit pair across different runs, but not across $\CZ$~gates during the same run.
	We randomize~$\leakcondphase[flux]$ and~$\leakcondphase[stat]$ as they have not been characterized in experiment and we instead capture an average behavior.

	Some potential errors are found to be small from the full-trajectory simulations of the $\CZ$~gate and thus are not included in the parametrized error model.
	The population exchange between $\ket{01}\leftrightarrow\ket{10}$, with coupling~$\coupling$, is suppressed~($<0.5\%$) since this avoided crossing is off-resonant by one anharmonicity~$\anharm$ with respect to~$\freq_{\interact}$.
	While $\ket{12}\leftrightarrow\ket{21}$ is also off-resonant by~$\anharm$, the coupling between these two levels is stronger by a factor of~2, hence potentially leading to a larger population exchange (see~\cref{sec:dynamics_leakage_subspace}).
	The $\ket{11}\leftrightarrow\ket{20}$~crossing is $2\anharm$~away from~$\freq_{\interact}$ and it thus does not give any substantial phase accumulation or population exchange~($<0.1\%$).
	We have compared the average gate fidelity of $\CZ$~gates simulated with the two methods and found differences below~$\pm 0.1\%$, demonstrating the accuracy of the parametrized model.

	\subsection{Effect of leakage on the code performance}
	\label{sub:code_performance}

	\begin{figure}
		\centering
		\includegraphics[width=\columnwidth]{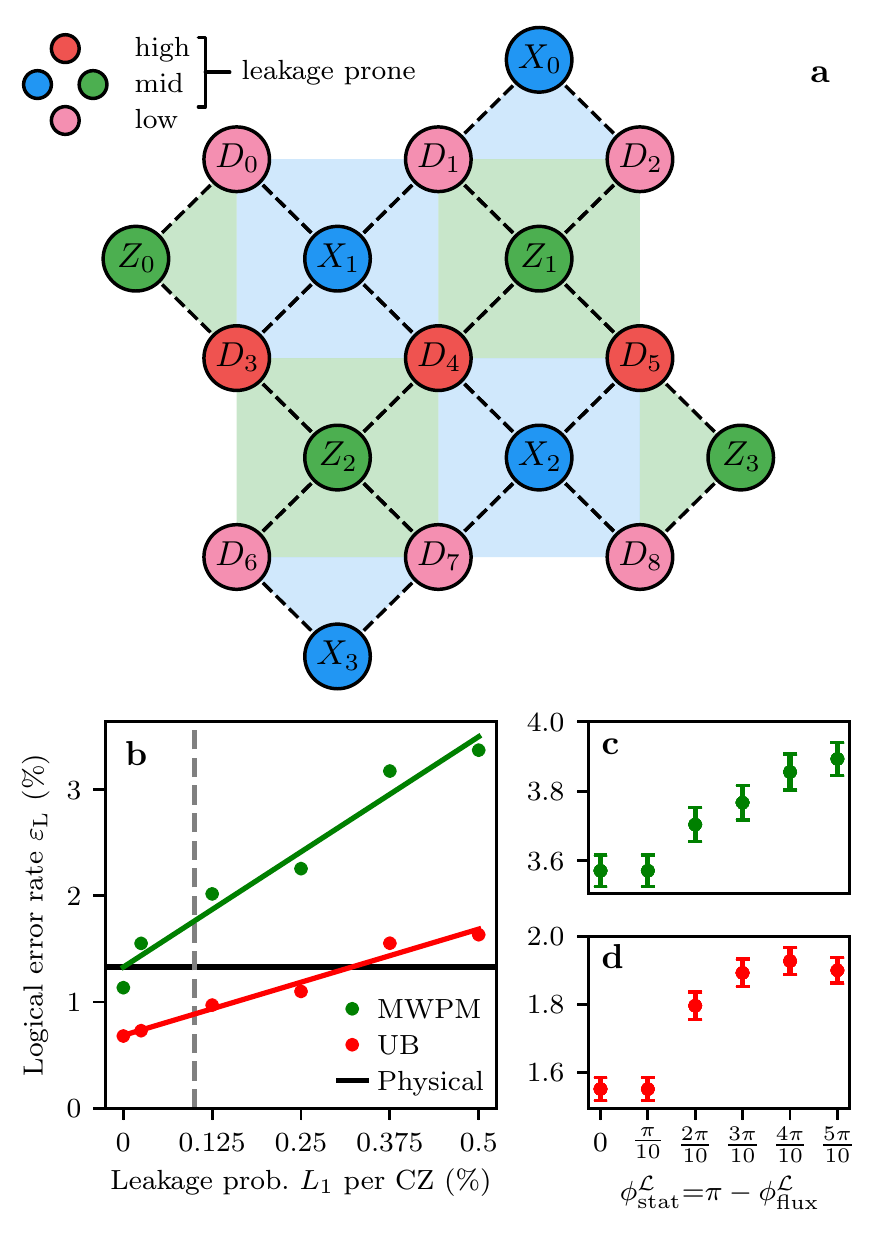}
		\caption{\label{fig:figure2_s17}
			\textbf{a}~Schematic overview of the Surface-17 layout~\cite{Versluis17}.
			Pink~(resp.~red) circles with \(\datatype\)~labels represent low- (high-) frequency data qubits, while blue (resp.~green) circles with \(\xtype\)~(\(\ztype\))~labels represent ancilla qubits of intermediate frequency, performing \(\xtype\)-type~(\(\ztype\)-type) parity checks.
			\textbf{b}~Dependence of the logical error rate~\(\logerrrate\) on the leakage probability~\(\leakrate\) for a MWPM~decoder (green) and for the decoding upper bound (red).
			The black solid line shows the physical error rate of a single transmon qubit.
			The dashed line corresponds to the recently achieved~\(\leakrate\) in experiment~\cite{Rol19a}.
			Logical error rate~\(\logerrrate\) for MWPM~(\textbf{c}) and upper bound~(\textbf{d}) as a function of the leakage conditional phases~$\leakcondphase[flux]$ and~$\leakcondphase[stat]$ (for~\(\leakrate=0.5\%\)).
			Here, these phases are not randomized but fixed to the given values across all runs.
			The logical error rates are extracted from an exponential fit of the logical fidelity over 20~QEC~cycles and averaged over~\(10^{5}\) runs for~\textbf{b} and~\(2\times10^{4}\) runs for~\textbf{c,d}.
			Error bars correspond to~\(95\%\) confidence intervals estimated by bootstrapping (not included in~\textbf{b} due the error bars being smaller than the symbol size).
		}
	\end{figure}

	We implement density-matrix simulations~\cite{quantumsim_website} to study the effect of leakage in Surface-17~(\cref{fig:figure2_s17}).
	We follow the frequency arrangement and operation scheduling proposed in~\cite{Versluis17}, which employs three qubit frequencies for the surface-code lattice, arranged as shown in~\cref{fig:figure2_s17}~\textbf{a}.
	The $\CZ$~gates are performed between the high-mid and mid-low qubit pairs, with the higher frequency qubit of the pair taking the role of~\(\qubit[\flux]\) (see~\cref{fig:figure1_CZmodel}).
	Based on the leakage model in~\cref{sub:leakage_error_model}, only the high and mid frequency qubits are prone to leakage (assuming no leakage mobility).
	Thus, in the simulation those qubits are included as three-level systems, while the low-frequency ones are kept as qubits.
	Ancilla-qubit measurements are modeled as projective in the \(\setbrkt{\ket{0}, \ket{1},\ket{2}}\)~basis and ancilla qubits are not reset between QEC~cycles.
	As a consequence, given the ancilla-qubit measurement~\(\meas[\cycle]\) at QEC~cycle~\(\cycle\), the syndrome is given by \(\meas[\cycle]\oplus\meas[\cycle-1]\) and the surface-code defect~\(\defect[\cycle]\) by \(\defect[\cycle]=\meas[\cycle]\oplus\meas[\cycle-2]\).
	For the computation of the syndrome and defect bits we assume that a measurement outcome~\(\meas[\cycle]=2\) is declared as~\(\meas[\cycle]=1\).
	The occurrence of an error is signaled by~\(\defect[\cycle]=1\).
	To pair defects we use a minimum-weight perfect-matching~(MWPM) decoder, whose weights are trained on simulated data without leakage~\cite{Obrien17,Spitz17} and we benchmark its logical performance in the presence of leakage errors.
	The logical qubit is initialized in~\(\ket{0}_{\logical}\) and the logical fidelity is calculated at each QEC~cycle, from which the logical error rate~\(\logerrrate\) can be extracted~\cite{Obrien17}.

	\Cref{fig:figure2_s17}~\textbf{b} shows that the logical error rate~$\logerrrate$ is sharply pushed above the memory break-even point by leakage.
	We compare the MWPM~decoder to the decoding upper bound~(UB), which uses the complete density-matrix information to infer a logical error.
	A strong increase in~$\logerrrate$ is observed for this decoder as well.
	Furthermore, the logical error rate has a dependence on the leakage conditional phases for both decoders, as shown in~\cref{fig:figure2_s17}~\textbf{c,d}.

	\subsection{Projection and signatures of leakage}
	\label{sub:projection_signatures}

	\begin{figure}[htp!]
		\centering
		\includegraphics[width=\columnwidth]{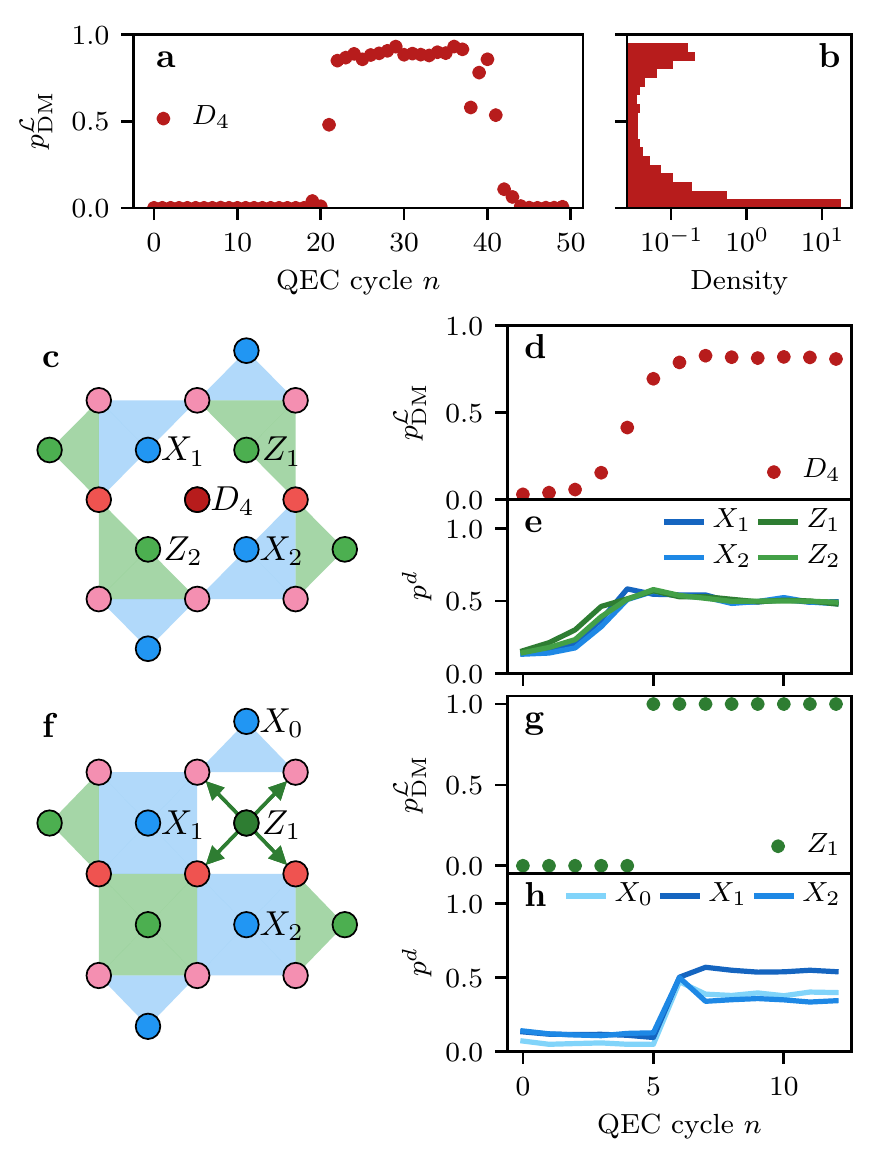}
		\caption{\label{fig:figure3_projection_and_signatures}
			Projection and signatures of leakage.
			\textbf{a}~Example realization of a data-qubit leakage event, extracted from the density-matrix simulations.
			\textbf{b}~Density histogram of all data-qubit leakage probabilities over 30~bins, extracted over $4\times10^{4}$~runs of 50~QEC~cycles each.
			\textbf{c-e}~Signatures of data-qubit leakage.
			\textbf{c}~Sketch of how leakage on a data qubit, e.g.~$\dataq{4}$, alters the interactions with neighboring stabilizers, leading to their anti-commutation (see~\cref{sec:anticommutation}).
			\textbf{d}~The average projection of the leakage probability \(\probleak[\DM]\) of \(\dataq{4}\) over all events, where this probability is first below and then above a threshold of~\(\probleak[\threshold]=0.5\) for at least 5 and 8~QEC~cycles, respectively.
			\textbf{e}~The average number of defects on the neighboring stabilizers of~$\dataq{4}$ over the selected rounds, showing a jump when leakage rises above~\(\probleak[\threshold]\).
			\textbf{f-h}~Signatures of ancilla-qubit leakage.
			\textbf{f}~Sketch of how leakage on an ancilla qubit, e.g.~$\zanc{1}$, effectively disables the stabilizer check and probabilistically introduces errors on the neighboring data qubits.
			\textbf{g}~We select realizations where~$\zanc{1}$ was in the computational subspace for at least 5~QEC~cycles, after which it was projected into~$\ket{2}$ by the readout and remained in that state for at least 5~QEC~cycles.
			\textbf{h}~The corresponding defect rate on neighboring stabilizers during the period of leakage.
			The error bars, which were estimated by bootstrapping, are smaller than the symbol sizes.
		}
	\end{figure}

	We now characterize leakage in Surface-17 and the effect that a leaked qubit has on its neighboring qubits.
	From the density matrix~(DM), we extract the probability~\(\probleak[\DM]\brkt{\qubitind}=\mathbb{P}(\qubitind\in\leaksub)=\braket{2|\rho_{\qubitind}|2}\) of a qubit~\(\qubitind\) being in the leakage subspace~\(\leaksub\) at the end of a QEC~cycle, after the ancilla-qubit measurements, where~$\rho_{\qubitind}$ is the reduced density matrix of~$\qubitind$.

	In the case of data-qubit leakage, \(\probleak[\DM]\brkt{\qubitind}\)~sharply rises to values near unity, where it remains for a finite number of \QEC~cycles (on average~9~\QEC~cycles for the considered parameters, see~\cref{tab:sim_params}).
	An example showing this projective behavior (in the case of qubit~\(\dataq{4}\)), as observed from the density-matrix simulations, is reported in~\cref{fig:figure3_projection_and_signatures}~\textbf{a}.
	This is the typical behavior of leakage, as shown in~\cref{fig:figure3_projection_and_signatures}~\textbf{b} by the bi-modal density distribution of the probabilities~\(\probleak[\DM]\brkt{\qubitind}\) for all the high-frequency data qubits~$Q$.
	Given this projective behavior, we identify individual events by introducing a threshold~\(\probleak[\threshold]\brkt{\qubitind}\), above which a qubit is considered as leaked.
	Here we focus on leakage on~\(\dataq{4}\), sketched in~\cref{fig:figure3_projection_and_signatures}~\textbf{c}.
	Based on a threshold~\(\probleak[\threshold]\brkt{\dataq{4}}=0.5\), we select leakage events and extract the average dynamics shown in~\cref{fig:figure3_projection_and_signatures}~\textbf{d}.
	Leakage grows over roughly 3~QEC~cycles following a logistic function, reaching a maximum probability of approximately~0.8. We observe this behavior for all three high-frequency data qubits~$\dataq{3},\dataq{4},\dataq{5}$.

	We observe an increase in the defect probability of the neighboring ancilla qubits during data-qubit leakage.
	We extract the probability~\(\probdefect\) of observing a defect~\(\defect=1\) on the neighboring stabilizers during the selected data-qubit leakage events, as shown in~\cref{fig:figure3_projection_and_signatures}~\textbf{e}.
	As~\(\probleak[\DM]\brkt{\dataq{4}}\) reaches its maximum, \(\probdefect\)~goes to a constant value of approximately~0.5.
	This can be explained by data-qubit leakage reducing the stabilizer checks it is involved in to effective weight-3 anti-commuting checks, illustrated in~\cref{fig:figure3_projection_and_signatures}~\textbf{c} and as observed in~\cite{Bultink19}.
	This anti-commutation leads to some of the increase in~\(\logerrrate\) for the~MWPM and UB~decoders in~\cref{fig:figure2_s17}~\textbf{b}.
	Furthermore, we attribute the observed projection of leakage (see~\cref{fig:figure3_projection_and_signatures}~\textbf{d}) to a back-action effect of the measurements of the neighboring stabilizers, whose outcomes are nearly randomized when the qubit is leaked (see~\cref{sec:anticommutation}).
	The weight-3 checks can also be interpreted as gauge operators, whose pairwise product results in weight-6 stabilizer checks, which can be used for decoding~\cite{Auger17,Nagayama17,Stace10,Bravyi13}, effectively reducing the code distance from~3 to~2.

	We also find a local increase in the defect probability during ancilla-qubit leakage.
	For ancilla qubits, $\probleak[\DM]$~is defined as the leakage probability after the ancilla projection during measurement.
	Since in the simulations ancilla qubits are fully projected, $\probleak[\DM]\brkt{Q}=0,1$ for an ancilla qubit~$\qubitind$, allowing to directly obtain the individual leakage events, as shown in~\cref{fig:figure3_projection_and_signatures}~\textbf{g}.
	We note that an ancilla qubit remains leaked for 11~\QEC~cycles on average for the considered parameters (see~\cref{tab:sim_params}).
	We extract~\(\probdefect\) during the selected events, as shown in~\cref{fig:figure3_projection_and_signatures}~\textbf{h}.
	In the QEC~cycle after the ancilla qubit leaks, \(\probdefect\)~abruptly rises to a high constant value.
	We attribute this to the \(\ztype\)~rotations acquired by the neighboring data qubits during interactions with the leaked ancilla qubit, as sketched in~\cref{fig:figure3_projection_and_signatures}~\textbf{f} and described in~\cref{sub:leakage_error_model}.
	The angle of rotation is determined by~\(\leakcondphase[\flux]\) or~\(\leakcondphase[\static]\), depending on whether the leaked ancilla qubit takes the roles of~\(\qubit[\static]\) or~\(\qubit[\flux]\), respectively (see~\cref{sec:sim_protocol} for the scheduling of operations).
	In the case of \(\ztype\)-type parity checks, these phase errors are detected by the \(\xtype\)-type stabilizers.
	In the case of \(\xtype\)-type checks, the phase errors on data qubits are converted to bit-flip errors by the Hadamard gates applied on the data qubits, making them detectable by the \(\ztype\)-type stabilizers.
	Furthermore, while the ancilla qubit is leaked, the corresponding stabilizer measurement does not detect any errors on the neighboring data qubits, effectively disabling the stabilizer, as sketched in~\cref{fig:figure3_projection_and_signatures}~\textbf{f}.
	This, combined with the spread of errors, defines the signature of ancilla-qubit leakage and explains part of the observed increase in~\(\logerrrate\) for the~MWPM and UB~decoders in~\cref{fig:figure2_s17}~\textbf{b}.

	For both data and ancilla qubits, a leakage event is correlated with a local increase in the defect rate, albeit due to different mechanisms.
	However, interpreting the spread of defects as signatures of leakage suggests the possible inversion of the problem, allowing for effective leakage detection.

	\subsection{Hidden Markov Models}
	\label{sub:HMMs}
	
	\begin{figure}
		\centering
		\includegraphics[width=\columnwidth]{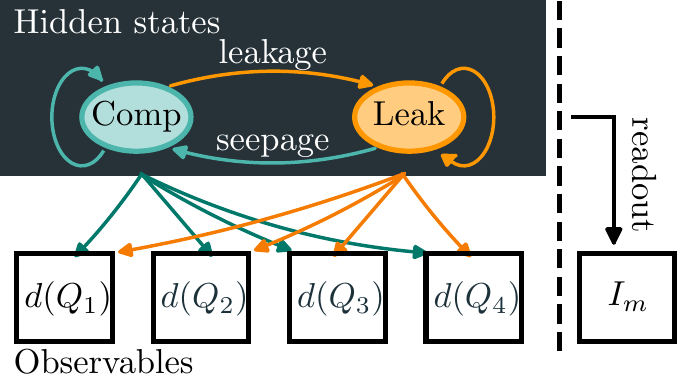}
		\caption{\label{fig:figure4_HMM_concept}
			Schematic representation of an~HMM for leakage detection.
			For both ancilla and data qubits only two hidden states are considered, corresponding to the qubit being either in the computational (teal) or leakage subspace (orange).
			Transitions between these states occur each QEC~cycle, depending on the leakage and seepage probabilities.
			The state-dependent observables are the defects~$\defect\brkt{\qubitind}$ on the neighboring stabilizers.
			For ancilla qubits, the in-phase component~\(\inphasecomp_{\meas}\) of the analog measurement is also used as an observable.
		}
	\end{figure}

	We use a set of~HMMs, one~HMM for each leakage-prone qubit, to detect leakage, similarly to what recently demonstrated in a minimal system~\cite{Bultink19}.
	An~HMM (see~\cref{fig:figure4_HMM_concept} and~\cref{sec:HMM_formalism}) models the time evolution of a discrete set of hidden states, the transitions between which are assumed to be Markovian.
	At each time step a set of observable bits is generated with state-dependent emission probabilities.
	Depending on the observed outcomes, the~HMM performs a Bayesian update of the predicted probability distribution over the hidden states.

	We apply the concept of~HMMs to leakage inference and outline their applicability for an accurate, scalable and run-time executable leakage-detection scheme.
	This is made possible by two observations.
	The first is that both data- and ancilla-qubit leakage are sharply projected~(see~\cref{sub:projection_signatures}) to high~\(\probleak[\DM]\brkt{\qubitind}\).
	This justifies the use of classical~HMMs with only two hidden states, corresponding to the qubit being in the computational or leakage subspace.

	The second observation is the sharp local increase in~\(\probdefect\) associated with leakage~(see~\cref{sub:projection_signatures}), which we identify as the signature of leakage.
	This allows us to consider only the defects on the neighboring stabilizers as relevant observables and to neglect correlations between pairs of defects associated with qubit errors.
	In the case of ancilla-qubit leakage, in addition to the defects, we consider the state information obtained from the analog measurement as input to the HMMs.
	Each transmon is dispersively coupled to a dedicated readout resonator.
	The state-dependent shift in the single-shot readout produces an output voltage signal, with in-phase and quadrature components (see~\cref{sec:meas_response_of_leakage}).
	
	The transition probabilities between the two hidden states are determined by the leakage and seepage probabilities per QEC~cycle, which are, to lowest order, a function only of the leakage probability~\(\leakrate\) per $\CZ$~gate and of the relaxation time~\(\Tone\) (see~\cref{sec:HMM_formalism}).
	We extract the state-dependent emission probabilities from simulation.
	When a qubit is not leaked, the probability of observing a defect on each of the neighboring stabilizers is determined by regular errors.
	When a data qubit is leaked, the defect probability is fixed to a nearly constant value by the stabilizer anti-commutation, while when an ancilla qubit is leaked, this probability depends on~\(\leakcondphase[\flux]\) and~\(\leakcondphase[\static]\).
	Furthermore, the analog measurement outcome can be used to extract a probability of the transmon being in~\(\ket{0},~\ket{1}\) or~\(\ket{2}\) using a calibrated measurement (see~\cref{sub:HMMs_ancillas} and~\cref{sec:meas_response_of_leakage}).

	\subsection{Data-qubit leakage detection}
	\label{sub:HMMs_data}

	\begin{figure}
		\centering
		\includegraphics[width=\columnwidth]{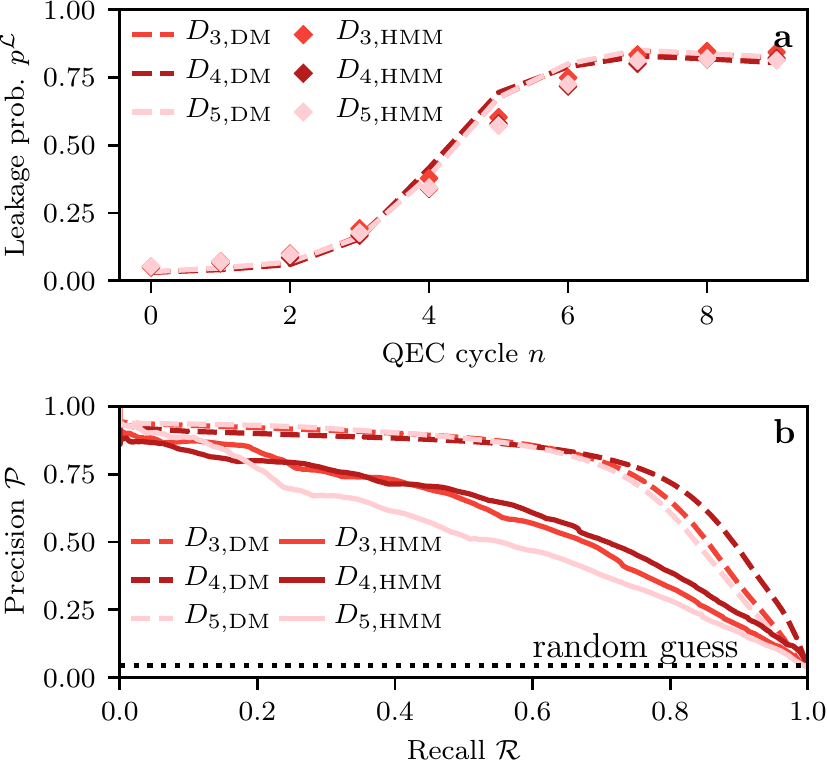}
		\caption{\label{fig:figure5_s17_data_response_precision_recall}
			\textbf{a}~Average response in time of the~HMMs (diamonds) to leakage, in comparison to the actual leakage probability extracted from the density-matrix simulations (dashed lines).
			The average is computed by selecting single realizations where~\(\probleak[\DM]\brkt{\qubitind}\) was below a threshold~\(\probleak[\threshold]=0.5\) for at least 5~QEC~cycles and then above it for~5 or more rounds.
			Error bars, estimated by bootstrapping, are smaller than the symbol sizes.
			\textbf{b}~Precision-recall curves for the data qubits over \(4\times10^{4}\)~runs of 50~\QEC~cycles each using the~HMM predictions (solid) and the leakage probability from the density matrix (dashed).
			The dotted line corresponds to a random guess classifier for which $\precision$~is equal to the fraction of leakage events (occurring with probability given by the density matrix) over all QEC~cycles and runs.
		}
	\end{figure}

	We assess the ability of the data-qubit~HMMs to accurately detect both the time and the location of a leakage event.
	The average temporal response~\(\probleak[\HMM]\brkt{\qubitind}\) of the~HMMs to an event is shown in~\cref{fig:figure5_s17_data_response_precision_recall} and compared to the leakage probabilities~\(\probleak[\DM]\brkt{\qubitind}\) extracted from the density-matrix simulation.
	Events are selected when crossing a threshold~\(\probleak[\threshold]\), as described in~\cref{sub:projection_signatures}, and the response is averaged over these events.
	For the data-qubit~HMMs, the response~\(\probleak[\HMM]\brkt{\qubitind}\) closely follows the probability~\(\probleak[\DM]\brkt{\qubitind}\) from the density matrix, reaching the same maximum leakage probability but with a smaller logistic growth rate.
	This slightly slower response is expected to translate to an average delay of about 1~QEC~cycles in the detection of leakage.

	We now explore the leakage-detection capability of the~HMMs.
	The precision~$\precision$ of an~HMM tracking leakage on a qubit~$\qubitind$ is defined as
	\begin{align}
	\precision[\HMM]\brkt{Q} = \prob\brkt{\qubitind \in \leaksub\mid\probleak[\HMM]\brkt{Q}>\probleak[\threshold]\brkt{Q}}
	\end{align}
	and can be computed as
	\begin{align}
	\precision[\HMM]\brkt{Q}=\frac{\sum_{i} \probleak[\DM]\brkt{\qubitind,i}\theta\bbrkt{\probleak[\HMM]\brkt{\qubitind,i}-\probleak[\threshold]\brkt{\qubitind}}}{\sum_{i} \theta\bbrkt{\probleak[\HMM]\brkt{\qubitind,i}-\probleak[\threshold]\brkt{\qubitind}}},
	\end{align}
	where $i$~runs over all runs and \QEC~cycles and $\theta$~is the Heaviside step function.
	The precision is then the fraction of correctly identified leakage events (occurring with probability given by the density matrix), over all of the HMM~detections of leakage.
	The recall~$\recall$ of an~HMM for a qubit~$\qubitind$ is defined as
	\begin{equation}
	\recall[\HMM]\brkt{\qubitind} = \prob\brkt{\probleak[\HMM]\brkt{Q} > \probleak[\threshold]\brkt{Q}\mid\qubitind \in \leaksub},
	\end{equation}
	and can be computed as
	\begin{align}
	\recall[\HMM]\brkt{Q}=\frac{\sum_{i} \probleak[\DM]\brkt{\qubitind,i}\theta\bbrkt{\probleak[\HMM]\brkt{\qubitind,i}-\probleak[\threshold]\brkt{\qubitind}}}{\sum_{i} \probleak[\DM]\brkt{\qubitind,i}}.
	\end{align}
	The recall is the fraction of detected leakage by the~HMM over all leakage events  (occurring with probability given by the density matrix).
	The precision-recall~(PR) of an~HMM (see~\cref{fig:figure5_s17_data_response_precision_recall}~\textbf{b}) is a parametric curve obtained by sweeping~\(\probleak[\threshold]\brkt{Q}\) and plotting the value of~$\precision$ and~$\recall$.
	Since the PR~curve is constructed from~\(\probleak[\HMM]\brkt{\qubitind}\) over all QEC~cycles and runs, it quantifies the detection ability in both time and space.
	The detection ability of an~HMM manifests itself as a shift of the PR~curve towards higher values of~\(\precision\) and~\(\recall\) simultaneously.
	We define the optimality~$\hmmopt\brkt{\qubitind}$ of the~HMM corresponding to qubit~$\qubitind$ as
	\begin{equation}
	\hmmopt\brkt{\qubitind}=\auc[\HMM]\brkt{\qubitind}/\auc[\DM]\brkt{\qubitind}, 
	\end{equation}
	where~\(\auc[\HMM]\brkt{\qubitind}\) is the area under the PR~curve of the \HMM~and~\(\auc[\DM]\brkt{\qubitind}\) is the area for the optimal model that predicts leakage with probability~$\probleak[\DM]\brkt{\qubitind}$, achieving the best possible~$\precision[\DM]$ and~$\recall[\DM]$.
	An average optimality of~\(\hmmopt\brkt{\qubitind}\approx74.0\%\) is extracted for the data-qubit~HMMs.
	Given the few QEC-cycle delay in the data-qubit HMM~response to leakage, the main limitation to the observed HMM~optimality~\(\hmmopt\brkt{\qubitind}\) is false detection when a neighboring qubit is leaked~(see~\cref{sec:more_HMM_analysis}).

	\subsection{Ancilla-qubit leakage detection}
	\label{sub:HMMs_ancillas}

	\begin{figure*}
		\centering
		\includegraphics[width=2\columnwidth]{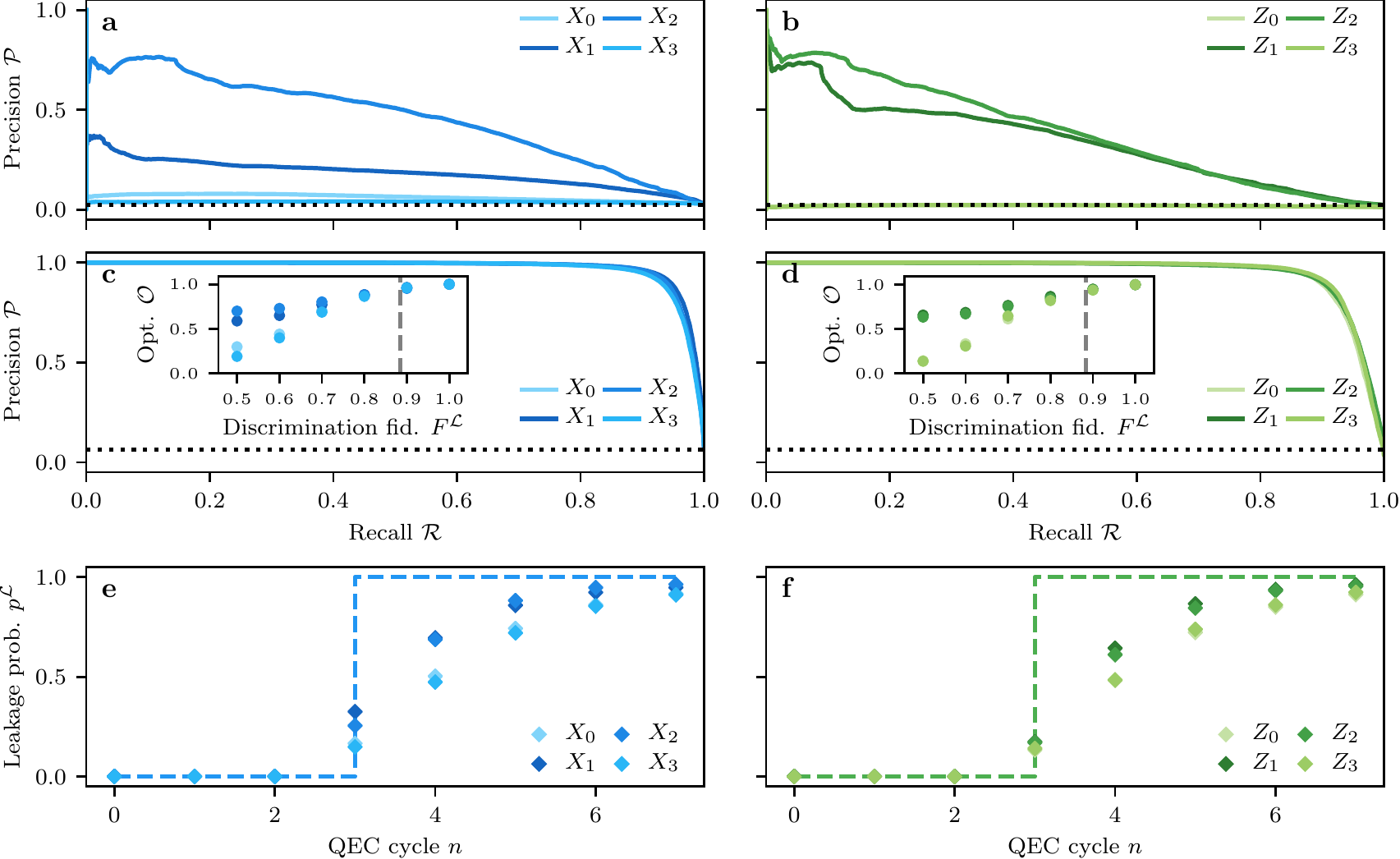}
		\caption{\label{fig:figure6_s17_anc_response_precision_recall}
			\textbf{a-d}~Precision-recall curves for the ancilla-qubit~HMMs over \(4\times10^{4}\)~runs of 50~\QEC~cycles each.
			In~\textbf{a,b} the~HMMs rely only on the observed defects on the neighboring stabilizers.
			In~\textbf{c-f} the~HMMs further get the in-phase component~$\inphasecomp[\meas]$ of the analog readout as input, from which~$\probleak[\meas]$ is extracted.
			The dotted line corresponds to a random guess classifier for which $\precision$~is equal to the fraction of leakage events over all QEC~cycles and runs.
			Insets in~\textbf{c,d}: the HMM optimality~\(\hmmopt\) as a function of the discrimination fidelity~\(\leakmeasfid\) between~\(\ket{1}\) and~\(\ket{2}\).
			The corresponding error bars (extracted over \(2\times10^{4}\)~runs of 20~\QEC~cycles each) are smaller than the symbol size.
			The vertical dashed line corresponds to the experimentally measured~$\leakmeasfid=88.4\%$.
			\textbf{e,f}~Average response in time of the ancilla-qubit HMMs (diamonds) to leakage, in comparison to the actual leakage probability extracted directly from the readout (dashed), extracted over \(4\times10^{4}\)~runs of 50~\QEC~cycles each.
			The average is computed by selecting single realizations where the qubit was in the computational subspace for at least 3~QEC~cycles and then in the leakage subspace for~5 or more.
		}
	\end{figure*}

	We now assess the ability of the ancilla-qubit~HMMs to accurately detect both the time and the location of a leakage event.
	These~HMMs take as observables the defects on the neighboring stabilizers at each QEC~cycle as well as the analog measurement outcome of the ancilla qubit itself.

	We first consider the case when the~HMMs rely only on the increase in the defect probability~\(\probdefect\) and show their PR~curves in~\cref{fig:figure6_s17_anc_response_precision_recall}~\textbf{a,b}.
	Bulk ancilla qubits have a low~\(\hmmopt\brkt{\qubitind}\approx35\%\), while boundary ancilla qubits possess no ability to detect leakage at all.
	We attribute this to the boundary ancilla qubits having only a single neighboring stabilizer, compared to bulk ancilla qubits having 3 in Surface-17.
	The~HMMs corresponding to pairs of same-type~($\xtype$ or~$\ztype$) bulk ancilla qubits exhibit significantly different PR~curves (see~\cref{fig:figure6_s17_anc_response_precision_recall}~\textbf{a,b}), despite the apparent symmetry of Surface-17.
	This symmetry is broken by the use of high- and low-frequency transmons as data qubits, leading to differences in the order in which an ancilla qubit interacts with its neighboring data qubits (see~\cite{Versluis17} and~\cref{fig:qec_circuit}), together with the fact that $\CZ$s with~$\leakrate\neq 0$ do not commute in general.
	In addition to a low~\(\hmmopt\brkt{\qubitind}\), the errors propagated by the leaked ancilla qubits (and hence the signatures of ancilla-qubit leakage) depend on~\(\leakcondphase[stat]\) and~\(\leakcondphase[flux]\) (randomized in the simulations).
	The values of these phases generally lead to different~\(\probdefect\) than the ones parameterizing the~HMM.
	The latter is extracted based on the average~\(\probdefect\) observed over the runs (see~\cref{sec:HMM_formalism}).
	In the worst-case (for leakage detection), these phases can lead to no errors being propagated onto the neighboring data qubits, resulting in the undetectability of leakage.
	The mismatch between the fluctuating~\(\probdefect\) (over~\(\leakcondphase[stat]\) and~\(\leakcondphase[flux]\)) and the average value hinders the ability of the ancilla-qubit~HMMs to detect leakage.
	Even if these phases were individually controllable, tuning them to maximize the detection capability of the~HMMs would also lead to an undesirable increase in~\(\logerrrate\) of a (leakage-unaware) decoder (see~\cref{fig:figure2_s17}).
	
	To alleviate these issues, we consider the state-dependent information obtained from the analog measurement outcome.
	The discrimination fidelity between~\(\ket{1}\) and~\(\ket{2}\) is defined as
	\begin{align}
	\leakmeasfid = 1 - \frac{\prob\brkt{1\mid2} + \prob\brkt{2\mid1}}{2},
	\end{align}
	where~\(\prob\brkt{i\mid j}\) is the conditional probability of declaring the measurement outcome~\(i\) given that the qubit has been prepared in state~\(\ket{j}\), assuming that no excitation or relaxation occur during the measurement (accounted for in post-processing).
	Here we assume that $\prob\brkt{0\mid2}=\prob\brkt{2\mid0}=0$, as observed in experiment (see~\cref{fig:sup_in_phase_responses_cal}).
	We focus on the discrimination fidelity as in our simulations relaxation is already accounted for in the measurement outcomes (see~\cref{sec:sim_protocol}).
	We extract~\(\leakmeasfid\) from recent experimental data~\cite{Bultink19}, where the readout pulse was only optimized to discriminate between the computational states.
	By taking the in-phase component of the analog measurement, for each state~\(\ket{j}\) a Gaussian distribution~\(\gaussian[j]\) is obtained, from which we get~\(\leakmeasfid=88.4\%\) (see~\cref{sec:meas_response_of_leakage}).
	
	In order to emulate the analog measurement in simulation, given an ancilla-qubit measurement outcome~\(\meas\in \setbrkt{0,1,2}\), we sample the in-phase response~\(\inphasecomp[\meas]\) from the corresponding distribution~\(\gaussian[\meas]\).
	The probability of the ancilla qubit being leaked given~$\inphasecomp_{\meas}$ is computed as
	\begin{align}
	\probleak[\meas] = \frac{\gaussian[2]\brkt{\inphasecomp[\meas]}}{\sum_{j\in\set{0,1,2}}\gaussian[j]\brkt{\inphasecomp[\meas]}}.
	\end{align}
	The ancilla-qubit~HMMs use the sampled responses~\(\inphasecomp[\meas]\), in combination with the observed defects, to detect leakage.

	The PR~curves of the~HMMs using the analog readout are shown in~\cref{fig:figure6_s17_anc_response_precision_recall}~\textbf{c,d}, from which an average~\(\hmmopt\brkt{\qubitind}\approx96\%\) can be extracted for the ancilla-qubit~HMMs.
	Given that projective measurements are used in the simulations, \(\auc[\DM]\brkt{\qubitind}=1\) for ancilla qubits.
	The temporal responses of the~HMMs to leakage are shown in~\cref{fig:figure6_s17_anc_response_precision_recall}~\textbf{e,f}, showing a sharp response to a leakage event, with an expected delay in the detection of at most 2~QEC~cycles.
	A further benefit of the inclusion of the analog-measurement information is that the detection capability of the~HMMs is now largely insensitive to the fluctuations in~\(\leakcondphase[\static]\) and~\(\leakcondphase[\flux]\).

	We explore~\(\hmmopt\brkt{\qubitind}\) as a function of~\(\leakmeasfid\), as shown in the inset of~\cref{fig:figure6_s17_anc_response_precision_recall}~\textbf{c,d}.
	To do this, we model~\(\gaussian[j]\) for each state as symmetric and having the same standard deviation, in which case~\(\leakmeasfid\) is a function of their signal-to-noise ratio only (see~\cref{sec:meas_response_of_leakage}).
	At low~\(\leakmeasfid\)~\(\brkt{\lesssim60\%}\) the detection of leakage is possible but very limited, especially for the boundary ancilla qubits.
	On the other hand, even at moderate values of~\(\leakmeasfid\)~\(\brkt{\approx 80\%}\), corresponding to experimentally achievable values, ancilla-qubit leakage can be accurately identified for both bulk and boundary ancilla qubits.
	Furthermore, relying solely on the analog measurements would allow for the potential minimization of the error spread associated with ancilla-qubit leakage, given controllability over~\(\leakcondphase[\static]\) and~\(\leakcondphase[\flux]\), without compromising the capability of the~HMMs to detect leakage.

	\subsection{Improving code performance via post-selection}\label{sub:restoring_code_performance}
	\label{sub:code_restoration}
	
	\begin{figure}
		\centering
		\includegraphics[width=\columnwidth]{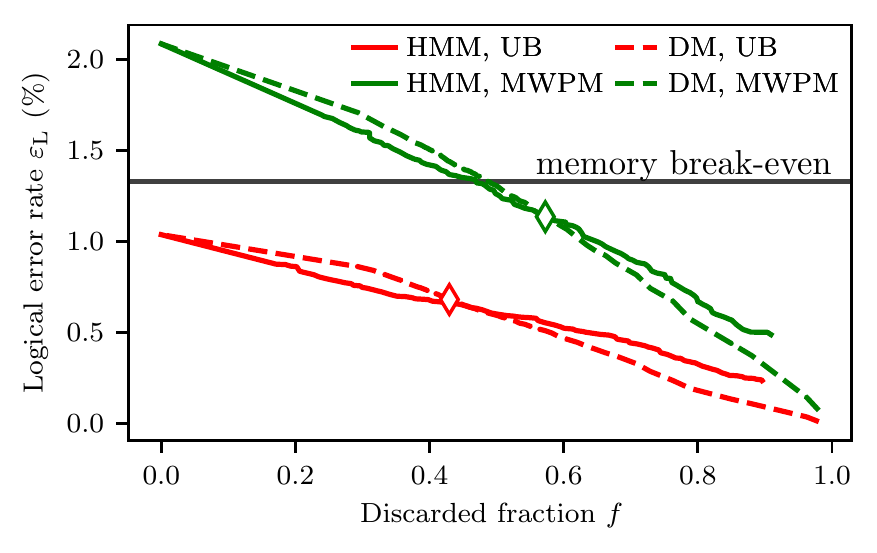}
		\caption{\label{fig:figure7_leak_mitigation_vs_discard}
			Improvement in the logical error rate~\(\logerrrate\) via post-selecting on the detection of leakage for a MWPM~decoder (green) and the decoder upper bound (red).
			The post-selection is based on the probabilities predicted by the~HMMs (solid) or on those extracted from the density-matrix simulation (dashed), for $2\times 10^4$~runs of 20~QEC~cycles each.
			The physical error rate of a single transmon qubit under decoherence is also given (solid black).
			Detection of leakage allows for the restoration of the performance of the MWPM~decoder, reaching the memory break-even point by discarding about~\(\approx47\%\) of the data.
			The logical error rates obtained from simulations without leakage (and without post-selection) are indicated by diamonds.
		}
	\end{figure}
	
	We use the detection of leakage to reduce the logical error rate~\(\logerrrate\) via post-selection on leakage detection, with the selection criterion defined as
	\begin{align}
	\label{eqn:discard_criteria}
	\max_{\qubitind,\cycle}\probleak\brkt{\qubitind,\cycle}\geq\probleak[\threshold]\brkt{\qubitind}.
	\end{align}
	We thus post-select any run for which the leakage probability of any qubit exceeds the defined threshold in any of the QEC~cycles.	
	Note that, while this criterion is insensitive to overestimation of the leakage probability due to a leaked neighboring qubit (see~\cref{sec:more_HMM_analysis}), it is sensitive to the correct detection of leakage in the first place and to the HMM~response in time (especially for short-lived leakage events).
	
	We perform the multi-objective optimization
	\begin{subequations}
		\begin{align*}
		&\underset{\probleak[\threshold]\brkt{\qubitind}}{\text{min}}
		&&\left(\datafrac,\logerrrate\right),\\
		&\text{subject to}
		&& 0.02 \leq \probleak[\threshold]\brkt{\qubitind} \leq 1,
		\end{align*}
	\end{subequations}
	where~\(\datafrac\) is the fraction of discarded data.
	The inequality constraint on the feasible space is helpful for the fitting procedure, required to estimate~\(\logerrrate\).
	This optimization uses an evolutionary algorithm~(\emph{NGSA-II}), suitable for conflicting objectives, for which the outcome is the set of lowest possible~\(\logerrrate\) for a given~\(\datafrac\).
	This set is known as the Pareto front and is shown in~\cref{fig:figure7_leak_mitigation_vs_discard} for both the~MWPM and UB~decoders.
	In~\cref{fig:figure7_leak_mitigation_vs_discard} we also compare post-selection based on the~HMMs against post-selection based on the density-matrix simulation.
	Here we use the predictions of the~HMMs which include the analog measurement outcome with the experimentally extracted~\(\leakmeasfid\) (see~\cref{sub:HMMs_ancillas}).
	The observed agreement between the two post-selection methods proves that the performance gain is due to discarding runs with leakage instead of runs with only regular errors.
	The performance of the MWPM~decoder is restored below the quantum memory break-even point by discarding~\(\datafrac\approx47\%\).
	The logical error rates extracted from simulations without leakage are achieved by post-selection of~\(\datafrac\approx57\%\) and~\(\datafrac\approx43\%\) of the data for the~MWPM and UB~decoders, respectively, when leakage is included.

	\section{Discussion}
	\label{sec:discussion}

	We have investigated the effects of leakage and its detectability using density-matrix simulations of a transmon-based implementation of Surface-17.
	Data and ancilla qubits tend to be sharply projected onto the leakage subspace, either indirectly by a back-action effect of stabilizer measurements for data qubits or by the measurement itself for ancilla qubits.
	During leakage, a large, but local, increase in the defect rate of neighboring qubits is observed.
	For data qubits we attribute this to the anti-commutation of the involved stabilizer checks, while for ancilla qubits we find that it is due to an interaction-dependent spread of errors to the neighboring qubits.
	We have developed a low-cost and scalable approach based on~HMMs, which use the observed signatures together with the analog measurements of the ancilla qubits to accurately detect the time and location of leakage events.
	The HMM~predictions are used to post-select out leakage, allowing for the restoration of the performance of the logical qubit below the memory break-even point by discarding~\(\datafrac\approx47\%\), opening the prospect of near-term QEC~demonstrations even in the absence of a dedicate leakage-reduction mechanism.
	
	A few noise sources have not been included in the simulations.
	First, we have not included readout-declaration errors, corresponding to the declared measurement outcome being different from the state in which the ancilla qubit is projected by the measurement itself.
	These errors are expected to have an effect on the performance of the MWPM~decoder, as well as a small effect on the observed optimality of the~HMMs.
	We have also ignored any crosstalk effects, such as residual couplings, cross-driving or dephasing induced by measurements on other qubits.
	While the presence of these crosstalk mechanisms is expected to increase the error rate of the code,
	it is not expected to affect the HMM~leakage-detection capability.
	We have assumed measurements to be perfectly projective.
	However, for small deviations, we do not expect a significant effect on the projection of leakage and on the observation of the characteristic signatures.
	
	We now discuss the applicability of~HMMs to other quantum-computing platforms subject to leakage and determine a set of conditions under which leakage can be efficiently detected.
	First, we assume single- and two-qubit gates to have low leakage probabilities, otherwise QEC~would not be possible in general.
	In this way, single- and two-qubit leakage probabilities can be treated as perturbations to block-diagonal gates, with one block for the computational subspace~$\compsub$ and one for the leakage subspace~$\leaksub$.
	We focus on the gates used in the surface code, i.e.,~$\CZ$ and Hadamard~$H$ (or $R_{Y}(\pi/2)$~rotations or equivalent gates).
	We consider data-qubit leakage first.
	We have observed that it is made detectable by the leakage-induced anti-commutation of neighboring stabilizers.
	The only condition ensuring this anti-commutation is that $H$~acts as the identity in~$\leaksub$ or that it commutes with the action of~$\CZ$ within the leakage block (see~\cref{sec:anticommutation}), regardless of the specifics of such action.
	Thus, data-qubit leakage is detectable via~HMMs if this condition is satisfied.
	In particular, it is automatically satisfied if~$\leaksub$ is 1-dimensional.
	We now consider ancilla-qubit leakage.
	Clearly, ancilla-qubit leakage detection is possible if the readout discriminates computational and leakage states perfectly or with high fidelity.
	If this is not the case, the required condition is that leaked ancilla qubits spread errors according to non-trivial leakage conditional phases, constituting signatures that can be used by an~HMM.
	If even a limited-fidelity readout is available, it can still be used to strengthen this signal, as demonstrated in~\cref{sub:HMMs_ancillas}.
	An issue is the possibility of the readout to project onto a superposition of computational and leakage subspaces.
	In that case, the significance of ancilla-qubit leakage is even unclear.
	However, for non-trivial leakage conditional phases, we expect a projection effect to the leakage subspace by a back-action of the stabilizer measurements, due to leakage-induced errors being detected onto other qubits, similarly to what observed for data qubits.

	The capability to detect the time and location of a leakage event demonstrated by the~HMMs could be used in conjunction with leakage-reductions units~(LRUs)~\cite{Aliferis07}.
	These are of fundamental importance for fault tolerance in the presence of leakage, since in~\cite{Suchara15} a threshold for the surface code was not found if dedicated~LRUs are not used to reduce the leakage lifetime beyond the one set by the relaxation time.
	While the latter constitutes a natural~LRU by itself, we do not expect it to ensure a threshold since, together with a reduction in the leakage lifetime, it leads to an increase in the regular errors due to relaxation.
	A few options for~LRUs in superconducting qubits are the swap scheme introduced in~\cite{Ghosh13_B}, or the use of the readout resonator to reset a leaked data-qubit into the computational subspace, similarly to~\cite{Egger18,Magnard18}.
	An alternative is to use the $\ket{02}\leftrightarrow\ket{11}$~crossing to realize a \quotes{leakage-reversal} gate that exchanges the leakage population in~$\ket{02}$ to~$\ket{11}$.
	An even simpler gate would be a single-qubit $\pi$~pulse targeting the~$\ket{1}\leftrightarrow\ket{2}$ transition.
	All these schemes introduce a considerable overhead either in hardware (swap, readout resonator), or time (swap, readout resonator, leakage-reversal gate), or they produce leakage when they are applied in the absence of it (leakage-reversal gate, $\pi$~pulse).
	Thus, all these schemes would benefit from the accurate identification of leakage, allowing for their targeted application, reducing the average circuit depth and minimizing the probability of inadvertently inducing leakage.

	We discuss how decoders might benefit from the detection of leakage.
	Modifications to MWPM~decoders have been developed for the case when ancilla-qubit leakage is directly measured~\cite{Suchara15,Kelly15}, and when data-qubit leakage is measured in the LRU~circuits~\cite{Suchara15}.
	Further decoder modifications might be developed to achieve a lower logical error rate relative to a leakage-unaware decoder, by taking into account the detected leakage and the probability of leakage-induced errors, as well as the stabilizer information that can still be extracted from the superchecks (see~\cref{sec:anticommutation}).
	In the latter case, a decoder could switch back and forth from standard surface-code decoding to e.g.~the partial subsystem-code decoding in~\cite{Stace10,Nagayama17,Auger17}.
	Given control of the leakage conditional phases, the performance of this decoder can be optimized by setting~\(\leakcondphase[\static]=\pi\) and~\(\leakcondphase[\flux]=0\), minimizing the spread of phase errors on the neighboring data qubits by a leaked ancilla qubit, as well as the noise on the weight-6 stabilizer extraction in the case of a leaked data qubit (see~\cref{sec:anticommutation}).
	Given a moderate discrimination fidelity of the leaked state, this is not expected to compromise the detectability of leakage, as discussed in~\cref{sub:HMMs_ancillas}.
	At the same time, for such a decoder we expect the improvement in the logical error rate to be limited in the case of low-distance codes such as Surface-17, as single-qubit errors can result in a logical error.
	This is because leakage effectively reduces the code distance, either because a leaked data qubit is effectively removed from the code, or because of the fact that a leaked ancilla qubit is effectively disabled and in addition spreads errors onto neighboring data qubits.
	Large codes, for which leakage could be well tolerated (depending on the distribution of leakage events), cannot be studied with density-matrix simulations, as done in this work for Surface-17.
	However, the observed sharp projection of leakage and the probabilistic spread of errors justify the stochastic treatment of this error~\cite{Suchara15}.
	Under the assumption that amplitude and phase damping can be modeled stochastically as well, we expect that the performance of decoders and LRUs in large surface codes can be well approximated in the presence of leakage.
	
	\begin{acknowledgments}
		We thank M.A.~Rol for useful discussions and C.C.~Bultink for providing us 			readout data from experiment.
		B.M.V., F.B.~and B.M.Te.~are supported by ERC grant EQEC No.~682726, B.M.Ta., V.P.O.~and L.D.C.~by the Office of the Director of National Intelligence (ODNI) and Intelligence Advanced Research Projects Activity (IARPA), via the U.S.~Army Research Office grant W911NF-16-1-0071, T.E.O.~by the Netherlands Organization for Scientific Research (NWO/OCW) under the NanoFront and StartImpuls programs, and by Shell Global Solutions BV.
		The views and conclusions contained herein are those of the authors and should not be interpreted as necessarily representing the official policies or endorsements, either expressed or implied, of the ODNI, IARPA, or the U.S.~Government.
	\end{acknowledgments}

	\appendix
	
	\section{Simulation protocol}
	\label{sec:sim_protocol}
	
	\begin{figure}
		\centering
		\includegraphics[width=\columnwidth]{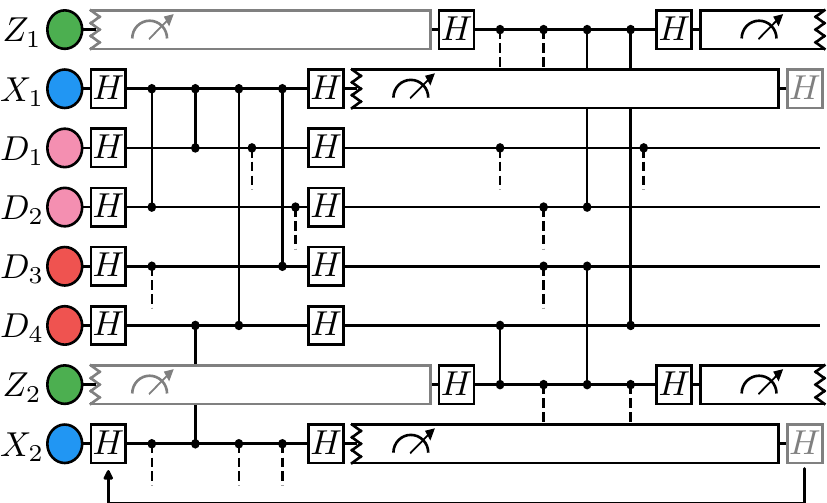}
		\caption{\label{fig:qec_circuit}
			The quantum circuit for a single \QEC~cycle employed in simulation, for the unit-cell scheduling defined in~\cite{Versluis17}.
			The qubit labels and frequencies correspond to the lattice arrangement shown in~\cref{fig:figure2_s17}.
			Gray elements correspond to operations belonging to the previous or the following \QEC~cycle.
			The \(\xtype\)-type parity checks are performed at the start of the cycle, while the \(\ztype\)-type parity checks are executed immediately after the \(\ztype\)-type stabilizer measurements from the previous cycle are completed.
			The duration of each operation is given in~\cref{tab:sim_params}.
			The arrow at the bottom indicates the repetition of \QEC~cycles.
		}
	\end{figure}
	
	For the Surface-17 simulations we use the open-source density-matrix simulation package \textit{quantumsim}~\cite{Obrien17}, available at~\cite{quantumsim_website}.
	For decoding we use a MWPM~decoder~\cite{Obrien17}, for which the weights of the possible error pairings are extracted from Surface-17 simulations via adaptive estimation~\cite{Spitz18} without leakage~(\(\leakrate=0\)) and an otherwise identical error model (described in~\cref{sec:sim_parameters}).
	
	The logical performance of the surface code as a quantum memory is the ability to maintain a logical state over a number of QEC~cycles.
	We focus on the \(\ztype\)-basis logical~\(\logstate{0}\), but we have observed nearly identical performance for~\(\logstate{1}\).
	We have not performed simulations for the \(\xtype\)-basis logical states \(\logstate{\pm}=\frac{1}{\sqrt{2}}\left(\logstate{0} \pm \logstate{1}\right)\), as previous studies did not observe a significant difference between the two bases~\cite{Obrien17}.
	The state~\(\logstate{0}\) is prepared by initializing all data qubits in~\(\ket{0}\), after which it is maintained for a fixed number of QEC~cycles (maximum~20 or~50 in this work), with the quantum circuit given in~\cref{fig:qec_circuit}.
	The first QEC~cycle projects the logical qubit into a simultaneous eigenstate of the \(\xtype\)- and \(\ztype\)-type stabilizers~\cite{Fowler12}, with the \(\ztype\)~measurement outcomes being~\(+1\) in the absence of errors, while the \(\xtype\)~outcomes are random.
	The information about the occurred errors is provided by the stabilizer measurement outcomes from each QEC~cycle, as well as by a \(\ztype\)-type stabilizer measurements obtained by measuring the data qubits in the computational basis at the end of the run.
	This information is provided to the MWPM~decoder, which estimates the logical state at the end of the experiment by tracking the Pauli frame.
	For decoding, we assume that the \(\ket{2}\)~state is  measured as a~\(\ket{1}\), as in most current experiments.
	In~\cref{sub:HMMs_ancillas} we considered the discrimination of~\(\ket{2}\) in readout, which can be used for leakage detection.
	While this information can be also useful for decoding, we do not consider a leakage-aware decoder in this work.
	
	The logical fidelity~\(\logfid[\cycle]\) at a final QEC~cycle~\(\cycle\) is defined as the probability that the decoder guess for the final logical state matches the initially prepared one.
	The logical error rate~\(\logerrrate\) is extracted by fitting the decay as
	\begin{align}
	\logfid[\cycle]=\frac{1}{2}\left[1+\brkt{1-2\logerrrate}^{\cycle-\cycle_{0}}\right],
	\end{align}
	where~\(\cycle_{0}\) is a fitting parameter (usually close to~0)~\cite{Obrien17}.
	
	\section{Error model and parameters}
	\label{sec:sim_parameters}
	
	\begin{table}
		\begin{tabular}{@{} *5l @{}}    \toprule
			Parameter & Value &  \\ \midrule
			\hline \hline
			Relaxation time \(\Tone\)  & 30~\(\us\) \\
			\hline
			Sweetspot dephasing time~\(\tdephsweet\) & 60~\(\us\) \\
			\hline
			High-freq.~dephasing time \\ at interaction point~\(\tdephint\) & 8~\(\us\) \\
			\hline
			Mid-freq.~dephasing time  \\ at interaction point~\(\tdephint\) & 6~\(\us\) \\
			\hline
			Mid-freq.~dephasing time \\ at parking point~\(\tdephpark\) & 8~\(\us\) \\
			\hline
			Low-freq.~dephasing time \\ at parking point~\(\tdephpark\) & 9~\(\us\) \\
			\hline
			Single-qubit gate time~\(\tgate\) & 20~\(\ns\) \\
			\hline
			Two-qubit interaction time~\(\tint\) & 30~\(\ns\) \\
			\hline
			Single-qubit phase-correction time~\(\tphasecorr\) & 10~\(\ns\) \\
			\hline
			Measurement time~\(\tmeas\) & 600~\(\ns\) \\
			\hline
			QEC-cycle time~\(\tcycle\) & 800~\(\ns\) \\ \bottomrule
			\hline
		\end{tabular}
		\caption{\label{tab:sim_params}
			The parameters for the qubit decoherence times and for the gate, measurement and \QEC-cycle durations used in the density-matrix simulations.}
	\end{table}
	
	In the simulations we include qubit decoherence via amplitude-damping and phase-damping channels.
	The time evolution of a single qubit is given by the Lindblad equation
	\begin{align}
	\frac{d\rho}{d t} = -\imnum\com{\hamil,\dmsymbol} + \sum_{i}\lindbladop[i]\dmsymbol\lindbladop[i]^{\dagger} - \frac{1}{2}\anticom{\lindbladop[i]^{\dagger}\lindbladop[i],\dmsymbol},
	\end{align}
	where~\(\hamil\) is the transmon Hamiltonian
	\begin{align}
	\hamil = \freq a^\dagger a + \frac{\anharm}{2} (a^\dagger)^2 a^2,
	\end{align}
	with $a$ the annihilation operator, $\freq$ and $\anharm$ the qubit frequency and anharmonicity, respectively, and~\(\lindbladop[i]\) the Lindblad operators.
	Assuming weak anharmonicity, we model amplitude damping for a qutrit by
	\begin{align}
	\lindbladop[amp] = \sqrt{\frac{1}{\Tone}}a.
	\end{align}
	The \(\ket{2}\)~lifetime is then characterized by a relaxation time~\(\Tone/2\).
	Dephasing is described by
	\begin{align}
	\lindbladop[deph,1] = \sqrt{\frac{8}{9\Tdeph}}
	\begin{pmatrix}
	1 & 0 & 0 \\
	0 & 0 & 0 \\
	0 & 0 & -1
	\end{pmatrix},
	\end{align}
	\begin{align}
	\lindbladop[deph,2] = \sqrt{\frac{2}{9\Tdeph}}
	\begin{pmatrix}
	1 & 0 & 0 \\
	0 & -1 & 0 \\
	0 & 0 & 0
	\end{pmatrix},
	\end{align}
	\begin{align}
	\lindbladop[deph,3] = \sqrt{\frac{2}{9\Tdeph}}
	\begin{pmatrix}
	0 & 0 & 0 \\
	0 & 1 & 0 \\
	0 & 0 & -1
	\end{pmatrix},
	\end{align}
	leading to a dephasing time~\(\Tdeph\) between \(\ket{0}\)~(resp.~\(\ket{1}\)) and \(\ket{1}\)~(\(\ket{2}\)), and to a dephasing time~\(\Tdeph/2\) between~\(\ket{0}\) and~\(\ket{2}\)~\cite{Rol19a}.
	The Lindblad equation is integrated for a time~\(t\) to obtain an amplitude- and phase-damping superoperator~\(\ptmop_{\ampdamp,t}\), expressed in the Pauli Transfer Matrix representation.
	For a gate~\(\ptmop_{\gate}\) of duration~\(t_{\gate}\), decoherence is accounted by applying \(\ptmop_{\ampdamp, t_{\gate}/2}\ptmop_{\gate}\ptmop_{\ampdamp, t_{\gate}/2}\).
	For idling periods of duration~\(t_{\idle}\), \(\ptmop_{\ampdamp,t_{\idle}}\) is applied.
	
	For single-qubit gates we only include the amplitude and phase damping experienced over the duration~\(\tgate\) of the gate.
	These gates are assumed to not induce any leakage, motivated by the low leakage probabilities achieved~\cite{Chen16b,Motzoi09}, and to act trivially in the leakage subspace.
	For two-qubit gates, namely the~\(\CZ\), we further consider the increased dephasing rate experienced by qubits when fluxed away from their sweetspot.
	In superconducting qubits, flux noise shows a typical power spectral density~\(S_{f} = A/f\), where~\(f\) is the frequency and~\(\sqrt{A}\) is a constant.
	In this work we consider~\(\sqrt{A}=4~\mu\Phi_{0}\), where~\(\Phi_{0}\) is the flux quantum.
	Both low- and high-frequency components are contained in~\(S_{f}\), which we define relative to the $\CZ$~gate duration~\(\tcz\).
	Away from the sweetspot frequency~\(\freq_{\sweet}\), a flux-tunable transmon has first-order flux-noise sensitivity \(D_{\phi}=\frac{1}{2\pi}\left|\frac{\partial \freq}{\partial \Phi}\right|\).
	The high-frequency components are included as an increase in the dephasing rate~\(\Gamma_{\phi}=1/\Tdeph\) (compared to the sweetspot), given by \(\Gamma_{\phi}=2\pi\sqrt{\ln 2A}D_{\phi}\)~\cite{Luthi18}, while the low-frequency components are not included due to the built-in echo effect of Net-Zero pulses~\cite{Rol19a}.
	High- and mid-frequency qubits are fluxed away to different frequencies, with dephasing rates computed with the given formula.
	Furthermore, during a few gates low-frequency qubits are fluxed away to a \quotes{parking} frequency in order to avoid unwanted interactions~\cite{Versluis17}.
	The computed dephasing times at the interaction point are given in~\cref{tab:sim_params}.
	For the $\CZ$~gates, we include this increased dephasing during the time~\(\tint\) spent at the interaction point, while for the phase-correction pulses of duration~\(\tphasecorr\) we consider the same dephasing time as at the sweetspot.
	We do not include deviations in the ideal single-qubit phases of the $\CZ$~gate~\(\phi_{01}=0\) and~\(\phi_{10}=0\) and the two-qubit phase~\(\phi_{11}=\pi\), under the assumption that gates are well tuned and that the low-frequency components of the flux noise are echoed out~\cite{Rol19a}.
	
	We now consider the coherence of leakage in the \(\CZ\)~gates, which in the rotating frame of the qutrit is modeled as the exchanges
	\begin{align}
	\ket{11} &\mapsto \sqrt{1-4\leakrate}\ket{11} + e^{i\phi}\sqrt{4\leakrate}\ket{02}, \label{eq:CZ_11}\\
	\ket{02} &\mapsto \sqrt{1-4\leakrate}\ket{02} - e^{-i\phi}\sqrt{4\leakrate}\ket{11},
	\label{eq:CZ_02}
	\end{align}
	with~\(\leakrate\) the leakage probability~\cite{Wood18}.
	The phase~\(\phi\) can lead to an interference effect between consecutive applications of the $\CZ$~gate across pairs of data and ancilla qubits.
	In terms of the full density matrix, the dynamics of~\cref{eq:CZ_11,eq:CZ_02} leads to a coherent superposition of computational and leaked states
	\begin{align}
	\rho = \begin{pmatrix}
	\begin{array}{c|c}
	\rho^{\compsub} & \rho^{\mathrm{coh}} \\
	\hline
	\rho^{\mathrm{coh}} & \rho^{\leaksub}
	\end{array}
	\end{pmatrix},
	\end{align}
	where \(\rho^{\compsub}\)~(resp.~\(\rho^{\leaksub}\)) is the density matrix restricted to the computational (leakage) subspace, while~\(\rho^{\mathrm{coh}}\) are the off-diagonal elements between these subspaces.
	We observe that varying the phase~\(\phi\) does not have an effect on the dynamics of leakage or on the logical error rate.
	We attribute this to the fact that each ancilla qubit interacts with a given data qubit only once during a QEC~cycle and it is measured at the end of it (and as such it is dephased).
	Thus, the ancilla-qubit measurement between consecutive $\CZ$~gates between the same pair prevents any interference effect.
	Furthermore, setting~\(\rho^{\mathrm{coh}}=0\), does not affect the projection and signatures of leakage nor the logical error rate (at least for the logical state prepared in the \(\ztype\)~basis), leading to an incoherent leakage model.
	We attribute this to the projection of leakage itself, which leaves the qubit into a mostly incoherent mixture between the computational and leakage subspaces.
	In the harmonic rotating frame, \(\ket{2}\)~is expected to acquire an additional phase during periods of idling, proportional to the anharmonicity.
	However, following the reasoning presented above, we also believe that this phase is irrelevant.

	An incoherent leakage model offers significant computational advantage for density-matrix simulations.
	For the case where~\(\rho_{\mathrm{coh}}\neq0\), the size of the stored density matrix at any time is~\(4^{6}\times9^{4}\) (6~low-frequency data qubits, 3~high-frequency data qutrits plus 1~ancilla qutrit currently performing the parity check).
	Setting~\(\rho_{\mathrm{coh}}=0\) reduces the size of the density matrix to~\(4^{6}\times5^{4}\), since for each qutrit only the~\(\ket{2}\bra{2}\) matrix element is stored in addition to the computational subspace.
	Thus, for the simulations in this work we rely on an incoherent model of leakage.
	
	Measurements of duration~\(\tmeas\) are modeled by applying \(\ptmop_{\ampdamp,\tmeas/2}\ptmop_{\mathrm{proj}}\ptmop_{\ampdamp,\tmeas/2}\), where \(\ptmop_{\ampdamp,\tmeas/2}\) are periods of amplitude and phase damping and \(\ptmop_{\mathrm{proj}}\) is a projection operator.
	This projector is chosen according to the Born rule and leaves the ancilla qubit in either~\(\ket{0}\), \(\ket{1}\) or~\(\ket{2}\).
	We do not include any declaration errors, which are defined as the measurement outcome being different from the state of the ancilla qubit immediately after the projection.
	Furthermore, we do not include any measurement-induced leakage, any decrease in the relaxation time via the Purcell effect or any measurement-induced dephasing via broadband sources.
	We do not consider non-ideal projective measurements (leaving the ancilla in a superposition of the computational states) due to the increased size of the stored density matrix that this would lead to.
	
	\section{Transmon measurements in experiment}
	\label{sec:meas_response_of_leakage}
	
	\begin{figure}
		\centering
		\includegraphics[width=\columnwidth]{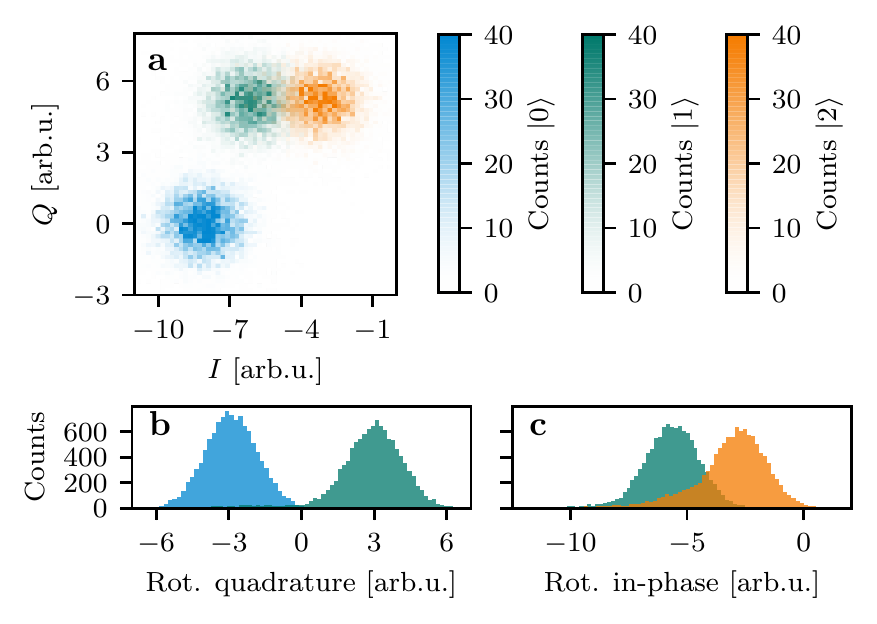}
		\caption{\label{fig:sup_in_phase_responses_cal}
			The analog measurement of transmons as extracted from experiment.
			\textbf{a}~Histograms of the in-phase~\(\inphasecomp\) and quadrature~\(\quadraturecomp\) components of the measured readout for a transmon prepared in~\(\ket{0}\), \(\ket{1}\) or~\(\ket{2}\).
			\textbf{b}~The histograms of the responses for the transmon initially prepared in~\(\ket{0}\) or~\(\ket{1}\), projected along the rotated quadrature maximizing the discrimination fidelity~\(\fid^{01}=99.6\%\).
			\textbf{c}~The histograms of the responses for the transmon initialized in~\(\ket{1}\) or~\(\ket{2}\), projected along the \(\inphasecomp\)~axis, in which case discrimination is achieved with a fidelity~\(\fid^{12}=88.4\%\).
		}
	\end{figure}
	
	We consider the measurements of transmons in experiment~\cite{Bultink19}, which is enabled by the dispersive coupling between a transmon and a dedicated readout resonator.
	The resonator is connected to a common feedline via a dedicated Purcell filter~\cite{Heinsoo18}.
	Measurement is performed by applying a readout pulse to the feedline, populating the resonator with photons.
	Each transmon induces a state-dependent shift of the frequency of the readout resonator, changing the amplitude and phase of the outgoing photons.
	This outgoing signal is amplified and the in-phase~(\(\inphasecomp\)) and quadrature~(\(\quadraturecomp\)) components are extracted.
	For calibration of the single-shot readout, the transmon is prepared in either~\(\ket{0}\), \(\ket{1}\) or~\(\ket{2}\) and subsequently measured.
	Repeating this experiment characterizes the spread of the~$\inphasecomp$ and~$\quadraturecomp$ components of each state~$\ket{i}$, which typically follow a two-dimensional Gaussian distribution~$\gaussian[i]$ with mean~\(\vec{\gausmean}_i\) and standard deviation~\(\vec{\gausstd}_i\) in the IQ~plane~\cite{Sank16,Heinsoo18}, as exemplified in~\cref{fig:sup_in_phase_responses_cal}~\textbf{a}.
	
	Given an analog measurement of~\(\inphasecomp\) and~\(\quadraturecomp\), the probability of a transmon being in state~\(\ket{i}\) can be expressed as
	\begin{align}
	\prob\brkt{i \mid \inphasecomp,\quadraturecomp} = \frac{\prob\brkt{\inphasecomp,\quadraturecomp \mid i}\prob\brkt{i}}{\prob\brkt{\inphasecomp,\quadraturecomp}},
	\end{align}
	where
	\begin{align}
	\prob\brkt{\inphasecomp,\quadraturecomp}=\sum_{j\in\set{0,1,2}}\prob\brkt{\inphasecomp,\quadraturecomp \mid j}\prob\brkt{j}.
	\end{align}
	We assume that the prior state probabilities are equally likely.
	Furthermore, given the typically observed Gaussian distributions, it holds that  \(\prob\brkt{\inphasecomp,\quadraturecomp \mid i} = \gaussian[i]\brkt{\inphasecomp,\quadraturecomp}\), which leads to
	\begin{align}
	\prob\brkt{i \mid \inphasecomp,\quadraturecomp} = \frac{\gaussian[i]\brkt{\inphasecomp,\quadraturecomp}}{\sum_{j\in\set{0,1,2}}\gaussian[j]\brkt{\inphasecomp,\quadraturecomp}}.
	\label{eq:prob_i_given_IQ}
	\end{align}
	
	In experiment, one is typically interested in discriminating between pairs of states~\(\ket{i}\) and~\(\ket{j}\), for which the discrimination fidelity is defined as
	\begin{align}
	\fid^{ij} = 1 - \prob\brkt{j \mid i}\prob\brkt{i} - \prob\brkt{i \mid j}\prob\brkt{j},
	\end{align}
	where~\(\prob\brkt{i \mid j}\) is the probability of declaring a measurement outcome~\(i\) given a prepared state~\(\ket{j}\), under the assumption of no excitation or relaxation during the measurement (accounted for in post-processing), and where~\(\prob\brkt{i}\) is the prior probability of the qubit being in state~\(\ket{i}\).
	Hence, the discrimination fidelity corresponds to the probability of correctly declaring the projected state.
	We focus on the discrimination fidelity as in our simulations relaxation is already accounted for in the measurement outcomes (see~\cref{sec:sim_protocol}).
	We assume \(\prob\brkt{i}=\prob\brkt{j}=\frac{1}{2}\), which leads to
	\begin{align}
	\fid^{ij} =1 - \frac{\prob\brkt{j \mid i} + \prob\brkt{i \mid j}}{2}.
	\end{align}
	This can be related to the signal-to-noise ratio~\(\SNR=\abs{\vec{\gausmean}_i-\vec{\gausmean}_j}/2\gausstd\), assuming symmetric Gaussian distributions, as
	\begin{align}
	\fid^{ij} = 1 - \frac{1}{2}\erfc\brkt{\frac{\SNR}{\sqrt{2}}}.
	\end{align}
	
	The IQ~response can be projected onto the axis joining the centers of a pair of two-dimensional Gaussian distributions, allowing to consider a single quadrature while maximizing the discrimination fidelity.
	Without loss of generality, we consider this optimal axis to be along~$\inphasecomp$.
	This allows to express~\cref{eq:prob_i_given_IQ} as
	\begin{align}
	\prob\brkt{i \mid \inphasecomp} = \frac{\gaussian[i]\brkt{\inphasecomp}}{\sum_{j\in\set{0,1,2}}\gaussian[j]\brkt{\inphasecomp}},
	\end{align}
	where~\(\gaussian[i]\brkt{\inphasecomp}\) is the marginal of~\(\gaussian[i]\brkt{\inphasecomp,\quadraturecomp}\).
	In experiment, in order to declare a binary measurement outcome, a threshold value for~$\inphasecomp$ is introduced, separating the regions for declaring either outcome.
	Following this approach, for a 3-outcome measurement, three projection axes are needed in general.
	However, since the Gaussian distributions for~$\ket{1}$ and~$\ket{2}$ are typically well-separated from the one for~$\ket{0}$, it is possible to use only two axes, i.e.,~one to discriminate~$\ket{0}$ from~$\ket{1}$, and one to further discriminate~$\ket{2}$ from the rest.
	For the measurement calibration from experiment~\cite{Bultink19}, shown in~\cref{fig:sup_in_phase_responses_cal}~\textbf{a}, the discrimination between~\(\ket{0}\) and~\(\ket{1}\) can be achieved by projecting the analog responses along a rotated quadrature axis which maximizes the discrimination fidelity~\(\fid^{01}=99.6\%\).
	Discriminating between~\(\ket{1}\) and~\(\ket{2}\) is performed with~\(\fid^{12}=88.4\%\) by projecting along a rotated in-phase axis, maximizing this fidelity.
	
	\section{Leakage-induced anti-commutation}
	\label{sec:anticommutation}
	
	\begin{figure}
		\centering
		\includegraphics[width=\columnwidth]{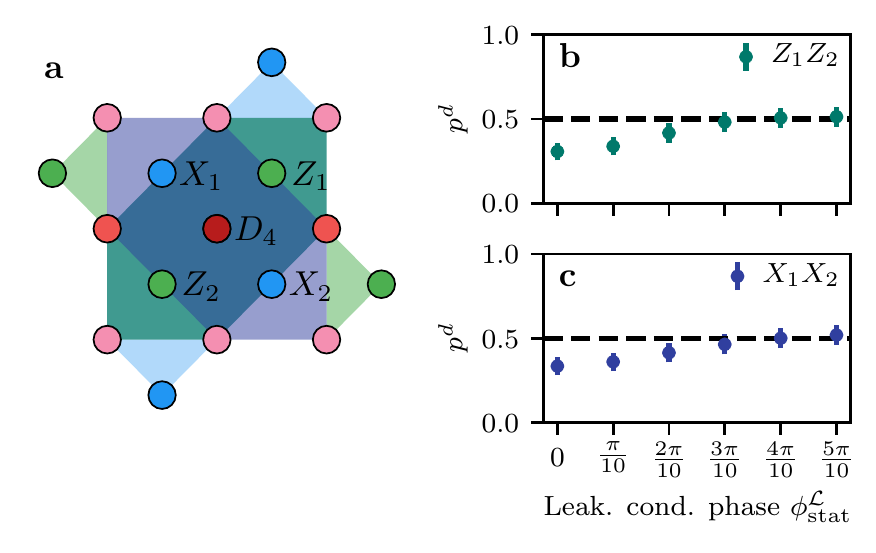}
		\caption{\label{fig:supercheck_signature}
			The effects of data-qubit leakage on the stabilizers of the code.
			\textbf{a}~Sketch of how data-qubit leakage in the bulk (e.g.~on~\(\dataq{4}\)) effectively defines weight-3 gauge operators, whose product forms a weight-6 \(\xtype\)-type~(purple) or \(\ztype\)-type~(teal) \quotes{supercheck} stabilizer, in addition to the standard weight-2 \(\xtype\)-type~(blue) and \(\ztype\)-type~(green) stabilizers.
			\textbf{b,c}~The average probability~\(\probdefect\) of observing a defect on the supercheck stabilizers during leakage on~\(\dataq{4}\) (defined by the leakage probability being above a threshold of~0.5) as a function of the leakage conditional phase~\(\leakcondphase[\static]\).
		}
	\end{figure}
	
	We study the behavior of neighboring stabilizers in the presence of a leaked data qubit.
	We focus on a parity-check operator in the bulk of the surface code.
	For the frequency scheme of~\cref{fig:figure2_s17}, this involves two leakage-prone high-frequency transmons and two low-frequency transmons, modeled as qutrits and qubits, respectively.
	The ancilla qubit used to perform the parity checks is leakage prone as well.
	However, here we do not consider this possibility, given the low probability of a pair of neighboring data and ancilla qubits to be leaked simultaneously.
	
	We consider the~$\CZ$ for transmons described in~\cref{sub:leakage_error_model}, without including any decoherence.
	In the limit of the leakage probability~$\leakrate\to 0$ (and leakage mobility~$\leakmobility\to 0$), for an ancilla qubit~$\anctype$ and a high-frequency data qubit~$\datatype$, the~$\CZ$ can be decomposed as
	\begin{align}
	&\ket{0}\bra{0}_\anctype \otimes \begin{pmatrix}
	1 & 0 & 0 \\
	0 & 1 & 0 \\
	0 & 0 & -1
	\end{pmatrix}_\datatype \nonumber\\
	&+\ket{1}\bra{1}_\anctype \otimes \begin{pmatrix}
	1 & 0 & 0 \\
	0 & -1 & 0 \\
	0 & 0 & -e^{-i\leakcondphase[\static]}
	\end{pmatrix}_\datatype \\
	&\eqqcolon \ket{0}\bra{0}_\anctype \otimes \leakyI_\datatype +\ket{1}\bra{1}_\anctype \otimes \leakyZ_\datatype.
	\label{eq:CZ_ideal_QHQM}
	\end{align}
	Note that~$\leakyI|_\compsub=\identity$ and~$\leakyZ|_\compsub=\ztype$, where~$\identity$ and~$\ztype$ are the standard identity and Pauli~$\ztype$ operators, respectively, and $\compsub$~is the qubit computational subspace.
	For a~$\CZ$ between an ancilla qubit and a low-frequency data qubit, it simply holds $\ket{0}\bra{0}_\anctype \otimes \identity_\datatype +\ket{1}\bra{1}_\anctype \otimes \ztype_\datatype$.
	Small values of~$\leakrate$, as observed in experiment~\cite{Rol19a}, can be treated as a perturbation to this.
	
	For a parity-check measurement, the back-action on the state of the data qubits is given by either one of two operators, depending on the outcome.
	In the case of a $\ztype$-type parity-check unit, these operators are given by
	\begin{align}
	M_\pm^\ztype = \frac{\leakyI_{abcd} \pm \leakyZ_{abcd}}{2},
	\end{align}
	where $\leakyI_{abcd}\coloneqq\leakyI_a\leakyI_b\identity_c\identity_d$ and $\leakyZ_{abcd}\coloneqq\leakyZ_a\leakyZ_b\ztype_c\ztype_d$.
	Under the assumption that single-qubit gates, namely the Hadamard gate, do not induce any leakage and act trivially on the leakage subspace, for the $\xtype$-type parity-check unit these operators are instead given by
	\begin{align}
	M_\pm^\xtype = \frac{\leakyI_{abcd} \pm \leakyX_{abcd}}{2},
	\end{align}
	where $\leakyX_{abcd}\coloneqq\leakyX_a\leakyX_b\xtype_c\xtype_d$ and
	\begin{align}
	\leakyX = \begin{pmatrix}
	0 & 1 & 0 \\
	1 & 0 & 0 \\
	0 & 0 & -e^{-i\leakcondphase[\static]}
	\end{pmatrix},
	\end{align}
	in which case $\leakyX|_\compsub=\xtype$ with~$\xtype$ the standard Pauli operator.
	
	The $\xtype$-type and $\ztype$-type parity checks commute if and only if~$M_\pm^\ztype$ and~$M_\pm^\xtype$ commute, as it holds
	\begin{align}
	\bbrkt{M_\pm^\ztype, M_\pm^\xtype} = \frac{1}{4} \bbrkt{\leakyZ_{abcd}, \leakyX_{abcd}}
	\end{align}
	(and also $\bbrkt{M_\pm^\ztype, M_\pm^\xtype} = -\bbrkt{M_\pm^\ztype, M_\mp^\xtype}$).
	To compute the commutator we first evaluate
	\begin{align}
	\bbrkt{\leakyZ, \leakyX} &=2i\begin{pmatrix}
	0 & -i & 0 \\
	i & 0 & 0 \\
	0 & 0 & 0
	\end{pmatrix},\\
	\anticom{\leakyZ, \leakyX}&=2\begin{pmatrix}
	0 & 0 & 0 \\
	0 & 0 & 0 \\
	0 & 0 & e^{-2i\leakcondphase[\static]}
	\end{pmatrix}.
	\end{align}
	It follows that
	\begin{align}
	\anticom{\leakyZ, \leakyX}\Big|_\compsub &= \anticom{Z,X}=0, \\
	\bbrkt{\leakyZ, \leakyX}\Big|_\leaksub&=0, \label{eq:XZcommute}
	\end{align}
	when restricted to the computational and leakage subspaces, respectively.
	Notice that, when all data qubits are in the computational subspace, it holds
	\begin{align}
	\anticom{\leakyZ_{abcd}, \leakyX_{abcd}}\Big|_{\compsub} = \anticom{\ztype_{abcd}, \xtype_{abcd}} = 0
	\end{align}
	as in the standard qubit case, where $\ztype_{abcd}=\ztype_a\ztype_b\ztype_c\ztype_d$ and $\xtype_{abcd}=\xtype_a\xtype_b\xtype_c\xtype_d$.
	Furthermore, we note that~\cref{eq:XZcommute} holds solely because~$H$ acts trivially in~$\mathcal{L}$ (as we assume) and it would continue to hold as long as~$H$ commutes with~$\CZ$ on~$\leaksub$.
	
	We now consider the case in which one of the high-frequency data qubits is in~$\leaksub$ (say~$a$) and the remaining ones are in~$\compsub$.
	In this case
	\begin{align}
	\anticom{\leakyZ_{abcd}, \leakyX_{abcd}}\Big|_{\leaksub_{a}} &= \anticom{-e^{-i\leakcondphase[\static]}\ztype_{bcd}, -e^{-i\leakcondphase[\static]}\xtype_{bcd}} \nonumber\\
	& = e^{-i\leakcondphase[\static]}\anticom{\ztype_{bcd}, \xtype_{bcd}}= 0.
	\label{eq:anticommXZleaked}
	\end{align}
	This shows that, in the presence of data-qubit leakage, $M_\pm^\ztype$ and~$M_\pm^\xtype$ do not commute.
	In particular, $\leakyZ_{abcd}$ and~$\leakyX_{abcd}$ anti-commute and this result is independent of the leakage conditional phase.
	Furthermore, it holds
	\begin{align}
	M_\pm^\ztype|_{\leaksub_a} = \frac{\identity_{bcd} \pm e^{-i\leakcondphase[\static]}\ztype_{bcd}}{2}
	\end{align}
	and similarly for $M_\pm^\xtype|_{\leaksub_a}$.
	
	For~$\leakcondphase[\static]=0,\pi$, $M_\pm^\xtype|_{\leaksub_a}$ are projectors onto the $\pm$-eigenspaces of~$\ztype_{bcd}$ or~$\xtype_{bcd}$, constituting effective weight-3 parity checks.
	In this case the anti-commutation~[\cref{eq:anticommXZleaked}] leads to fully randomized ancilla-qubit measurement outcomes, corresponding to a probability~$\probdefect=50\%$ of observing a defect each QEC~cycle on each of the neighboring stabilizers.
	However, the product of two weight-3 same-type checks is a weight-6 stabilizer of the surface code, thus the product of the two ancilla-qubit measurement outcomes corresponds to the parity of the 6~data qubits involved.
	In particular, the stabilizer group can be redefined as including the standard weight-4 checks which do not involve the leaked qubit, together with the defined weight-6 \quotes{superchecks}, while the weight-3 checks are gauge operators~\cite{Bravyi13,Stace10,Nagayama17,Auger17}, as illustrated in~\cref{fig:supercheck_signature}~\textbf{a}.
	For the superchecks to be correctly obtained, both $\xtype$-type gauge operators need to be measured before any of the two $\ztype$-type gauge operators (or viceversa), which already holds true for the circuit schedule we consider~\cite{Versluis17}.
	In the case of a leaked qubit on the boundary, only one supercheck operator can be defined (for a rotated surface code, this is a weight-4 $\xtype$- or $\ztype$-type boundary supercheck), while the other one must be ignored for decoding~\cite{Stace10,Nagayama17,Auger17}.
	In the case of one leaked data-qubit in Surface-17, the minimum weight of a dressed logical operator is~2, reducing the code distance by~1.
	For example, if~$\dataq{4}$ is leaked, two $\xtype$~errors on~$\dataq{2}$ and~$\dataq{7}$ constitute a logical~$\xtype$.
	In a larger surface code, the reduction of the distance depends on the number of leaked qubits, as well as their distribution on the lattice~\cite{Auger17}.
	
	In the general case where~$\leakcondphase[\static]\neq 0,\pi$, while the anti-commutation still holds, $M_\pm^\ztype|_{\leaksub_a}$ and~$M_\pm^\xtype|_{\leaksub_a}$ are not projectors and thus the ancilla-qubit measurement outcomes are not fully randomized, which is expected to have an effect on the observed~$\probdefect$.
	However, in the simulations $\probdefect\approx50\%$ for both the case when~$\leakcondphase[\static]$ is randomized across runs~(see~\cref{fig:figure3_projection_and_signatures}) or when it is fixed, independently of the specific value.
	Since the defects~$\defect$ are computed as $\defect[\cycle] = \meas[\cycle]\oplus \meas[\cycle-2]$, where~$\meas[\cycle]$ is the measurement outcome at QEC~cycle~$\cycle$, even a moderate imbalance between the probabilities of measuring~$\meas[\cycle] = 0$ and~$\meas[\cycle] = 1$ (fluctuating across QEC~cycles) can lead a defect probability~$\probdefect\approx 50\%$.
	Furthermore, the phase rotations depending on~$\leakcondphase[\static]$ affect the measurement of each of the two weight-3 gauge operators independently, which in turn undermines the correct extraction of the weight-6 stabilizer parity.
	This effect is observed in~\cref{fig:supercheck_signature}~\textbf{b,c}, where in the case of~$\leakcondphase[\static]=0,\pi$ the observed defect probability roughly corresponds to the expected one from a weight-6 check (relative to the observed one for the standard weight-4 and weight-2 checks in the absence of leakage), while a higher defect probability is observed otherwise, reaching up to~$50\%$.
	Hence, the control of~$\leakcondphase[\static]$ in experiment would be beneficial for decoding in the presence of data-qubit leakage whenever the superchecks are considered.

	\section{HMM formalism}
	\label{sec:HMM_formalism}

	An~HMM describes the time evolution of a set~\(S=\setbrkt{\state}\) of not directly observable states~\(\state\) (i.e.,~\quotes{hidden}), over a sequence of independent observables~\(\obs=\setbrkt{\obsi}\).
	At each time step~\(\cycle\) the states undergo a Markovian transition, such that the probability~\(\probstate\bbrkt{\cycle}\) of the system being in the state~$\state$ is determined by the previous distribution~\(\probstate\bbrkt{\cycle-1}\) over all~$\state\in S$.
	These transitions can be expressed via the transition matrix~\(A\), whose elements are the conditional probabilities \(A_{\state,\state'} \coloneqq \prob\brkt{\state\bbrkt{\cycle}=\state\mid \state\bbrkt{\cycle-1}=\state'}\).
	A set of observables is then generated with state-dependent probabilities~\(B_{\obsi\bbrkt{\cycle},\state} \coloneqq\prob\brkt{\obsi\bbrkt{\cycle}=\obsi\mid \state\bbrkt{\cycle}=\state}\).
	Inverting this problem, the inference of the posterior state probabilities~\(\probstate\bbrkt{\cycle}\) from the realized observables is possible via
	\begin{align}
	\probstate\bbrkt{\cycle}&= \prob\brkt{\state\bbrkt{\cycle}\mid\obs\bbrkt{\cycle},\obs\bbrkt{\cycle-1},\ldots,\obs\bbrkt{1}}\\
	&= \frac{\prob\brkt{\obs\bbrkt{\cycle}\mid \state\bbrkt{\cycle}}\probstate[\mathrm{prior}]\bbrkt{\cycle}}{\prob\brkt{\obs\bbrkt{\cycle}}} \\
	&= \frac{\prod_{i}\prob\brkt{\obsi\bbrkt{\cycle}\mid \state\bbrkt{\cycle}}\probstate[\mathrm{prior}]\bbrkt{\cycle}}{\prod_{i}\prob\brkt{\obsi\bbrkt{\cycle}}} \\
	&= \frac{\prod_{i}B_{\obsi\bbrkt{\cycle},\state}\probstate[\mathrm{prior}]\bbrkt{\cycle}}{\sum_{\state'}\prod_{i}B_{\obsi\bbrkt{\cycle},\state'}\probstateprime[\mathrm{prior}]\bbrkt{\cycle}},
	\end{align}
	where \(\probstate[\mathrm{prior}]\bbrkt{\cycle}\) is the prior probability
	\begin{align}
	\probstate[\mathrm{prior}]\bbrkt{\cycle}=\sum_{\state'}A_{\state,\state'}\probstateprime\bbrkt{\cycle-1}.
	\end{align}
	We define \(B_{\obs\bbrkt{\cycle},\state} = \prod_{i}B_{\obsi\bbrkt{\cycle},\state}\), which for discrete~\(\obsi\) constitute the entries of the emission matrix~\(B\).
	In addition to the transition and emission probabilities, the initial state probabilities \(\probstate\bbrkt{\cycle=0}\) are needed for the computation of the evolution.

	In the context of leakage detection, we consider only two hidden states, \(S=\setbrkt{\compsub,\leaksub}\), namely whether the qubit is in the computational~(\(\compsub\)) or the leakage subspace~(\(\leaksub\)).
	The transition matrix is parameterized in terms of the leakage and seepage probabilities per \QEC~cycle.
	The leakage probability is estimated as \(\Gamma_{\compsub\rightarrow \leaksub}\approx N_\flux\leakrate\) (for low~\(\leakrate\)), where~\(N_\flux\) is in how many $\CZ$~gates the qubit is fluxed during a \QEC~cycle and~\(\leakrate\) is the leakage probability per $\CZ$~gate.
	The seepage probability is estimated by \(\Gamma_{\leaksub \rightarrow \compsub}\approx N_\flux\seeprate + \brkt{1-e^{\frac{\tcycle}{\Tone/2}}}\), where~\(\tcycle\) is the \QEC~cycle duration and~$\Tone$ the relaxation time (see~\cref{tab:sim_params}), while~\(\seeprate\) is the seepage contribution from the gate, where~\(\seeprate=2\leakrate\) due to the dimensionality ratio between~\(\compsub\) and~\(\leaksub\) for a qubit-qutrit pair~\cite{Wood18}.
	The transition matrix~\(A\) is then given by
	\begin{align}
		A  = \begin{pmatrix}
			1 -\Gamma_{\compsub\rightarrow\leaksub} & \Gamma_{\leaksub\rightarrow\compsub} \\
			\Gamma_{\compsub\rightarrow\leaksub} & 1-\Gamma_{\leaksub\rightarrow\compsub}
		\end{pmatrix}.
	\end{align}
	We assume that all qubits are initialized in~\(\compsub\), which defines the initial state distribution~\(\probcomp\bbrkt{\cycle=0}=1\) used by the~HMMs.

	The~HMMs consider the defects~\(\defect\brkt{\qubitind[i]}\equiv\defect_i\) on the neighboring ancilla qubits \(\qubitind[i]\) at each \QEC~cycle, occurring with probability~\(p^{\defect_{i}}\), as the observables for leakage detection.
	Explicitly, the emission probabilities are parameterized in terms of the conditional probabilities~\(B_{\defect_i\bbrkt{n},s}=\prob\brkt{\defect_i\bbrkt{n}\mid s}\) of observing a defect when the modeled qubit is in~\(s=\compsub\) or~\(s=\leaksub\).
	We extract~\(B_{\defect_i\bbrkt{n},\compsub}\) directly from simulation, by averaging over all runs and all \QEC~cycles, motivated by the possible extraction of this probability in experiment.
	While this includes runs when the modeled qubit was leaked, we observe no significant differences in the HMM~performance when we instead post-select out these periods of leakage, which we attribute to the low~\(\leakrate\) per \(\CZ\)~gate.
	We extract~\(B_{\defect_i\bbrkt{n},\leaksub}\) from simulation over the \QEC~cycles when the leakage probability~\(\probleak[\DM]\brkt{\qubitind[i]}\) as observed from the density matrix is above a threshold of \(\probleak[\threshold]=0.5\).
	In the case of ancilla-qubit leakage, \(B_{\defect_i\bbrkt{n},\leaksub}\)~depends on the values of the leakage conditional phases~\(\leakcondphase[\static]\) and~\(\leakcondphase[\flux]\).
	Thus, in the case of randomized leakage conditional phases, the~HMMs are parameterized by the average~\(B_{\defect_i\bbrkt{n},\leaksub}\).
	In the case of data-qubit leakage, the extracted~\(B_{\defect_i\bbrkt{n},\leaksub}\) is~\(\approx0.5\) regardless of the leakage conditional phases, as expected from the anti-commuting stabilizers (see~\cref{sub:projection_signatures}).

	For ancilla-qubit leakage detection, the analog measurement outcome \(\inphasecomp[\meas]\) can be additionally considered as an observable, in which case \(\obs=\setbrkt{\defect_{i},\inphasecomp[\meas]}\).
	In this case, the state-dependent probability is further parametrized by \(B_{\inphasecomp[\meas]\bbrkt{n},\compsub}=\prob\brkt{\inphasecomp[\meas]\bbrkt{\cycle}\mid\compsub}=\gaussian[0]\brkt{\inphasecomp[\meas]\bbrkt{\cycle}}+\gaussian[1]\brkt{\inphasecomp[\meas]\bbrkt{\cycle}}\) and by \(B_{\inphasecomp[\meas]\bbrkt{n},\leaksub}=\prob\brkt{\inphasecomp[\meas]\bbrkt{\cycle}\mid\leaksub}=\gaussian[2]\brkt{\inphasecomp[\meas]\bbrkt{\cycle}}\), where \(\gaussian[i]\) are the Gaussian distributions of the analog responses in the IQ~plane, projected along a rotated in-phase axis~\(\inphasecomp\), following the same treatment as in~\cref{sec:meas_response_of_leakage}.

	\section{HMM error budget}
	\label{sec:more_HMM_analysis}

	\begin{figure}
		\centering
		\includegraphics[width=\columnwidth]{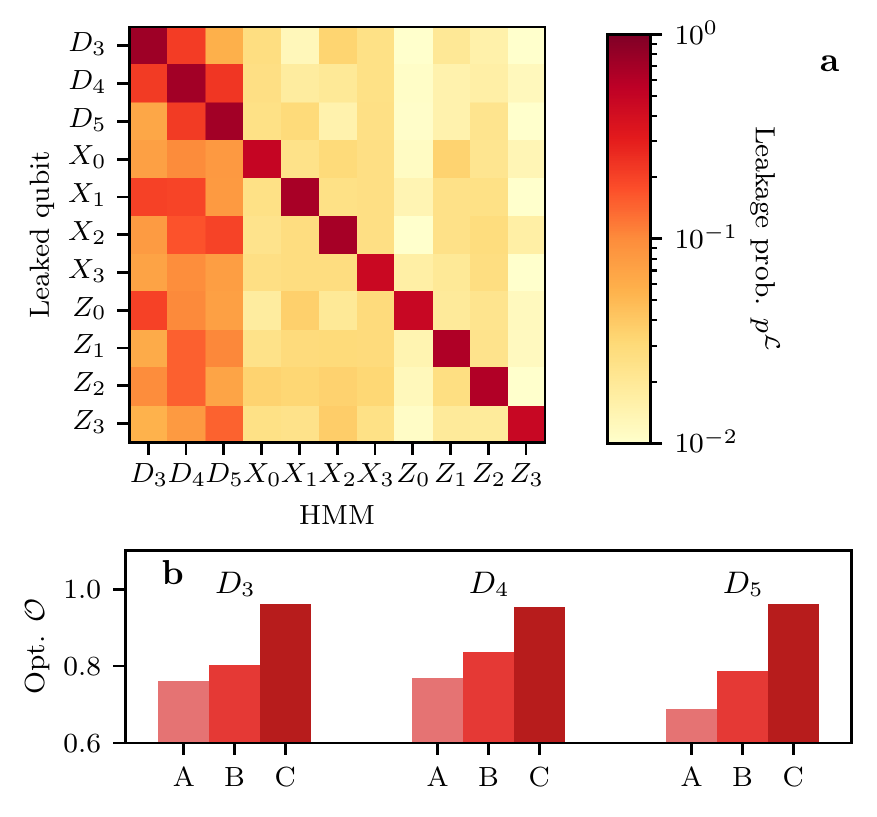}
		\caption{\label{fig:hmm_crosstalk_mat}
			The crosstalk between the~\HMMs.
			\textbf{a}~Average responses of all HMMs 1~\QEC~cycle after a given qubit leaks.
			We select individual realizations where the leakage probability~\(\probleak\) is first below and then above a threshold~\(\probleak[\threshold]=0.5\) for~5 and 8~\QEC~cycles, respectively.
			\textbf{b}~The extracted data-qubit HMM optimality~\(\hmmopt\).
			A:~optimality of the HMMs including all error sources.
			B:~runs where ancilla-qubit leakage was present (according to density matrix) are discarded.
			C:~leakage on any of the other data qubits (not tracked by the given~HMM) is discarded as well.
		}
	\end{figure}

	In this section we explore the limiting factors behind the remaining suboptimality of the HMMs presented in this work.
	The HMMs consider the probability of observing a defect at a given \QEC~cycle on each stabilizer independently, thus they do not take into account the correlations between defects due to regular errors.
	Data-qubit errors or hook errors (which are data-qubit errors propagated due to a single ancilla-qubit error during the parity-check circuit) give rise to a pair of correlated defects on different stabilizers either in the same \QEC~cycle or in two consecutive ones.
	Ancilla-qubit errors or measurement-declaration errors instead give rise to pairs of correlated defects on the same stabilizer and for one or two \QEC~cycles, respectively.
	As the HMMs take an increase in the defect probability as a signature of leakage, this is expected to result in the HMMs overestimating the probability of the tracked qubit being leaked.
	In addition, each HMM only takes the defects on the neighboring stabilizers as observables.
	Despite each HMM sharing observables with the neighboring ones, the probability of leakage at each \QEC~cycle is estimated independently by each HMM.
	While this choice minimizes the computational overhead, as a result each HMM is additionally prone to overestimating the probability of leakage when a neighboring qubit is leaked instead (leading to an increased defect probability observed on only a subset of the stabilizers taken as observables by the HMM).
	The HMMs can be expanded to account for these limitations, either by increasing the number of hidden states to model regular errors~\cite{Bultink19} or by expanding the set of observables to include next-nearest neighbor stabilizers, in order to account for leakage on neighboring qubits, in which case the HMMs would be still local and hence scalable.
	As either solution would increase the complexity and overhead of the models, we evaluate the contributions of each of these limitation to the detection capabilities of the HMMs.

	We first focus on the overestimation of the leakage probability predicted by the HMMs in the presence of leakage on a neighboring qubit, which we refer to as \quotes{HMM crosstalk}.
	We consider the detection scheme taking into account the analog measurements (with the currently achieved experimental discrimination fidelity~\(\leakmeasfid\), see~\cref{sub:HMMs_ancillas}).
	The average responses of all HMMs to leakage events on any qubit and the predicted leakage probability 1~\QEC~cycle after detection (defined by the predicted probability crossing a threshold of~0.5) are shown in~\cref{fig:hmm_crosstalk_mat}~\textbf{b}.
	The responses of the neighboring HMMs immediately (1-2~\QEC~cycles) after crossing this threshold is indicative of the likelihood of leakage being declared on a neighboring qubit (based on the extracted HMM responses shown in~\cref{fig:figure5_s17_data_response_precision_recall,fig:figure6_s17_anc_response_precision_recall}).
	Across individual runs, these parasitic responses can lead to false detections.
	Ancilla-qubit HMMs are insensitive to leakage on other data or ancilla qubits (see~\cref{fig:hmm_crosstalk_mat}).
	We attribute this to the use of the analog measurement outcomes which discriminate between~\(\ket{1}\) and~\(\ket{2}\) with moderate fidelity and between~\(\ket{0}\) and~\(\ket{2}\)) with very high fidelity.
	Instead, data-qubit HMMs are prone to overestimating the response in the case of leakage on other qubits.
	The crosstalk is proportional to the number of shared observables between the pairs of HMMs and depends on the expected defect probabilities during leakage by each model.

	We further break down the relative contributions to the optimality~\(\hmmopt\) (defined in~\cref{sub:HMMs_data}) of each of the data-qubit HMMs due to the crosstalk, shown in~\cref{fig:hmm_crosstalk_mat}~\textbf{b}.
	Post-selecting out runs where ancilla-qubit leakage is detected from the density matrix increases the average~\(\hmmopt\) of the three data-qubit HMMs from~\(\hmmopt\approx74.0\%\) to~\(\hmmopt\approx80.9\%\).
	Further post-selecting out events where leakage is detected on any of the other data qubits (which are not tracked by the given HMM) increases the average optimality to~\(\hmmopt\approx95.9\%\).
	The larger contribution from neighboring data-qubit leakage is consistent with the higher crosstalk~(see~\cref{fig:hmm_crosstalk_mat}~\textbf{a}) between data-qubit HMMs relative to the ancilla-qubit ones and constitutes the dominant limitation behind the HMM optimality.
	We attribute the remaining suboptimality to the presence of regular errors, caused by qubit relaxation and dephasing, and to the parametrization of the transition and output probabilities.

	\section{An alternative scheme for enhancing ancilla-qubit leakage detection}
	\label{sec:flagged_HMMs}
	
	We consider an alternative scheme (to the one considering the analog measurement outcomes) allowing for enhancing ancilla-qubit leakage detection beyond that achievable by only considering the increase in the defect probability on neighboring stabilizers.
	In this scheme a \(\pi\)~pulse is applied to each ancilla qubit every other \QEC~cycle, accounted for in post-processing.
	Under the assumption that a \(\pi\)~rotation has a trivial effect on a leaked qubit, the post-processed measurement outcomes (in the absence of errors) would show a flip every other \QEC~cycle during the period of leakage, which corresponds to a defect every \QEC~cycle.
	This scheme would require minimal overhead, as these rotations can be integrated with the existing single-qubit gates applied to the ancilla qubits at the start of each \QEC~cycle.
	A downside is that ancilla qubits would spend more time in the first excited state on average, increasing the effect of amplitude damping.
	We have not simulated this scheme, but we have investigated it entirely in post-processing by only applying flips to the measurement outcomes during periods of ancilla-qubit leakage (as extracted from the density matrix).
	Although this does not capture the increase in the ancilla-qubit error rate due to amplitude damping, we expect that it captures the effect of the scheme on the detection of leakage.
	
	The average HMM~optimality for the bulk~\(\xtype\) and~\(\ztype\) ancilla qubits is~\(\hmmopt\brkt{\xtype}\approx58.5\%\) and~\(\hmmopt\brkt{\ztype}\approx51.5\%\), respectively.
	For the boundary~\(\xtype\) and~\(\ztype\) ancilla qubits, it is~\(\hmmopt\brkt{\xtype}\approx74.0\%\) and~\(\hmmopt\brkt{\ztype}\approx43.0\%\), respectively.
	This constitutes a considerable increase in optimality relative to the scheme relying only on the observed defects (see~\cref{fig:figure6_s17_anc_response_precision_recall}~\textbf{a,b}).
	However, the artificially induced defects on leaked ancilla qubits lead to the increase in the crosstalk between ancilla- and data-qubit HMMs.
	This has the effect of lowering the average data-qubit HMM~optimality from~\(\hmmopt\brkt{\dataind}\approx74.0\%\) (see~\cref{sub:HMMs_data}) to~\(\hmmopt\brkt{\dataind}\approx39.8\%\).
	While such scheme may be beneficial for the post-selection-based scheme defined in~\cref{sub:code_restoration} (as in that case leakage detected on any qubits leads to discarding the run), it would be detrimental for leakage-aware decoding or targeted leakage-reduction units as these rely on the accurate detection in both time and space.
	
	\section{Second-order leakage effects}
	\label{sec:dynamics_leakage_subspace}
	
	\begin{figure}
		\centering
		\includegraphics[width=\columnwidth]{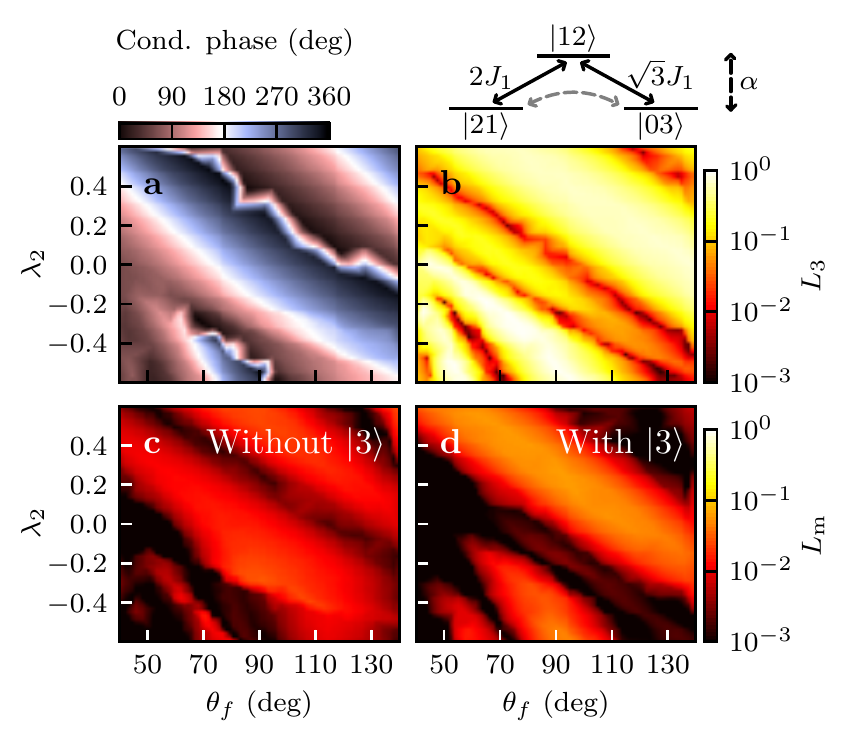}
		\caption{\label{fig:heatmaps}
			Heatmaps obtained from $\CZ$~full-trajectory simulations, including~(\textbf{a,b,d}) and not including~(\textbf{c})~$\ket{3}$ in the Hilbert space of the fluxing qubit.
			The conditional phase~$\phi_{\CZ}$~(\textbf{a}), $L_1$~(\textbf{b}) and~$\leakmobility$~(\textbf{c,d}) are plotted versus the flux-pulse parameters (see~\cite{Martinis14,Rol19a} for definitions).
			The interaction point is located at~$\theta_f=90~\degrees$.
			The inset (top-right) schematically shows the direct and effective couplings between levels in the 3-excitation manifold at the interaction point.
			The states $\ket{03}$ and~$\ket{21}$ are on resonance, while~$\ket{12}$ is detuned by one anharmonicity~$\anharm$.
		}
	\end{figure}
	
	\begin{table}
		\begin{tabular}{|c|cccccc|}
			\hline
			\multicolumn{1}{c}{} &~~~~  &~~~~ &~~~~~~  &~~~~ &~~~~ &~~~~ \\[\dimexpr-\normalbaselineskip-\arrayrulewidth]
			\textbf{Parameter} & \multicolumn{2}{c|}{$\dataq{low}$}& \multicolumn{2}{c|}{$\dataq{mid}$}& \multicolumn{2}{c|}{$\dataq{high}$}\\
			\hline \hline
			{$\freq/2\pi$ at sweet spot (GHz)} & \multicolumn{2}{c|}{4.9}& \multicolumn{2}{c|}{6.0}& \multicolumn{2}{c|}{6.7}\\
			\hline
			{$\anharm/2\pi$ (MHz)} & \multicolumn{2}{c|}{$-300$}& \multicolumn{2}{c|}{$-300$}& \multicolumn{2}{c|}{$-300$}\\
			\hline
			{$\coupling/2\pi$ at int.~point (MHz)} & \multicolumn{3}{c|}{15}& \multicolumn{3}{c|}{15}\\
			\hline
		\end{tabular}
		\caption{
			\label{tab:target_parameters}
			Parameters used in the~$\CZ$ full-trajectory simulations, with~$\anharm$ the anharmonicity and $\coupling$ the coupling.
			We follow the frequency scheme of~\cite{Versluis17} with the arrangement shown in~\cref{fig:figure2_s17}.
		}
	\end{table}
	
	In this section we consider exchanges between states in the leakage subspace as a result of a $\CZ$~gate acting on an already leaked qubit.
	We focus on the exchange between~$\ket{12}$ and~$\ket{21}$, referred to as \quotes{leakage mobility} in~\cref{sub:leakage_error_model}.
	We also expand the model to include~$\ket{3}$ on the fluxing qubit and consider the exchange between~$\ket{03}$ and~$\ket{12}$, which we call \quotes{superleakage}.
	
	The Hamiltonian of two transmons dispersively coupled via a bus resonator in the rotating-wave approximation is given by~\cite{Rol19a}
	\begin{align}
	H(t) =\:\, &\freq_{\static} a_{\static}^\dagger a_{\static} + \frac{\anharm_{\static}}{2} (a_{\static}^\dagger)^2 a_{\static}^2 \nonumber\\
	&+ \freq_{\flux}(\Phi(t))\, a_{\flux}^\dagger a_{\flux} + \frac{\anharm_{\flux}}{2} (a_{\flux}^\dagger)^2 a_{\flux}^2 \nonumber \\
	&+ \coupling(\Phi(t))\, (a_{\static} a_{\flux}^\dagger + a_{\static}^\dagger a_{\flux}),
	\label{eq:hamiltonian}
	\end{align}
	where $a$ is the annihilation operator, $\freq$ and~$\anharm$ are the qubit frequency and anharmonicity, respectively, and~$\coupling$ is the effective coupling mediated by virtual excitations through the bus resonator.
	We assume that this Hamiltonian is a valid approximation up to the included states.
	For this Hamiltonian, multiple avoided crossings are found when sweeping~$\freq_\flux$, as schematically shown in~\cref{fig:figure1_CZmodel}.
	We perform full-trajectory simulations (following the same structure as in~\cite{Rol19a}, excluding distortions and quasi-static flux noise) using the parameters reported in~\cref{tab:sim_params} and~\cref{tab:target_parameters}.
	Note that extending these simulations to~$\ket{3}$ does not affect the leakage probability~$\leakrate$ from the computational~($\compsub$) to the leakage subspace~($\leaksub$), nor the fidelity within~$\compsub$.
	
	We define the superleakage probability~$\superleak$ as
	\begin{equation}
	\superleak\coloneqq\abs{\braket{03|\mathcal{S}_{\CZ}(\ket{12}\bra{12})|03}}^2,
	\end{equation}
	where $\mathcal{S}_{\CZ}$~is the superoperator corresponding to the simulated noisy~$\CZ$.
	$\superleak$~can be high depending on the specific parameters of the flux pulse and of the system, as~\cref{fig:heatmaps}~\textbf{b} shows for the high-mid qubit pair, even when~$\phi_{\CZ}=\pi$ (see~\cref{fig:heatmaps}~\textbf{a}).
	We attribute this to the avoided crossing between $\ket{12}\leftrightarrow\ket{03}$ occurring at~$\freq_\interact+\abs{\anharm}$, where $\freq_\interact$ is the fluxing-qubit frequency at the interaction point.
	For fast-adiabatic flux pulses~\cite{Martinis14} (with respect to the $\ket{11}\leftrightarrow\ket{02}$ avoided crossing), pulsing the higher frequency qubit to the interaction point results in the near-diabatic passage through $\ket{12}\leftrightarrow\ket{03}$, inducing a Landau-Zener transition in which a small but finite population is transferred from~$\ket{12}$ to $\ket{03}$.
	At the $\CZ$~interaction point, the off-resonant interaction between~$\ket{12}$ and~$\ket{03}$ leads to a further population exchange, with a coupling strength~$\sqrt{3}\coupling$.
	Compared to the off-resonant exchange between~$\ket{01}$ and~$\ket{10}$, this interaction is stronger by a factor~$\sqrt{3}$, which can lead to large values of~$\superleak$ when combined with the initial population transfer to~$\ket{03}$ on the way to the avoided crossing.
	Furthermore, the phases acquired during the two halves of a Net-Zero pulse can lead to interference~\cite{Rol19a}, increasing or decreasing the $\ket{12}\leftrightarrow\ket{03}$ exchange population.
	Including the $\ket{12}\leftrightarrow\ket{03}$ crossing leads also to differences in the values of the leakage conditional phases.
	
	We now focus on leakage mobility, which occurs with probability~$\leakmobility$, defined as
	\begin{equation}
	\leakmobility\coloneqq\abs{\braket{21|\mathcal{S_{\CZ}}(\ket{12}\bra{12})|21}}^2.
	\end{equation}
	If~$\ket{3}$ is not included, $\leakmobility$~takes small but non-negligible values, as shown in~\cref{fig:heatmaps}~\textbf{c}.
	We attribute this to the off-resonant interaction between~$\ket{12}$ and~$\ket{21}$, with coupling strength~$2\coupling$.
	Even though this coupling is stronger than for $\ket{12}\leftrightarrow\ket{03}$, $\leakmobility$~is generally smaller than~$\superleak$ due to the fluxing qubit not passing through the $\ket{12}\leftrightarrow\ket{21}$ avoided crossing (located at $\freq_{\interact}-\abs{\anharm}$) on its way to the $\CZ$~interaction point.
	Including~$\ket{3}$, $\leakmobility$ can take higher values, as shown in~\cref{fig:heatmaps}, which we associate to a two-excitation exchange between~$\ket{03}$ and~$\ket{21}$, virtually mediated by~$\ket{12}$.
	While this is a two-excitation process, $\ket{21}$ and~$\ket{03}$ are on resonance at the interaction point, in which case the effective coupling can be estimated as the product of the bare couplings divided by the detuning with~$\ket{12}$, i.e.
	\begin{equation}
	\frac{1}{2\pi}\frac{(2\coupling)(\sqrt{3}\coupling)}{\anharm} \approx 2.6~\MHz,
	\end{equation}
	in analogy to the excitation exchange between a pair of transmons mediated virtually via the bus resonator.
	Since~$\ket{03}$ and~$\ket{21}$ are on resonance exactly at the interaction point only when $\anharm_{\flux}= \anharm_{\static}$, differences in the anharmonicities affect the strength of this exchange.
	
	\section{Effects of leakage mobility and superleakage on leakage detection and code performance}
	\label{sec:HMM_vs_leakmobility}
	
	We include leakage mobility in simulations,
	exploring the range of leakage-mobility probabilities~\(\leakmobility\in\left[0,1.5\%\right]\) for a fixed leakage probability~\(\leakrate=0.125\%\) and randomized leakage-conditional phases~\(\leakcondphase[\static]\) and~\(\leakcondphase[\flux]\) (see~\cref{sub:leakage_error_model}).
	Due to constraints imposed by the size of the density matrix, we only include leakage mobility between the high-frequency data qubits and the ancilla qubits.
	Thus, we have neglected the possibility of leakage being transferred to the low-frequency data qubits.
	
	Leakage mobility has a negligible effect on the logical performance of the code and the optimality of the HMMs.
	This is because leakage mobility is only significant in the case of an already leaked qubit, which occurs with a low probability across \QEC~cycles, given the low~\(\leakrate\) per \(\CZ\)~gate.
	Thus, the leakage swapping between neighboring qubits can be considered as a second-order effect and has a negligible impact on the logical error rate and HMM optimality extracted from the simulations.
	We also observe that the average duration of a leakage event on a given qubit is reduced in the presence of leakage mobility.
	
	We now consider the effect of superleakage (see~\cref{sec:dynamics_leakage_subspace}) on the logical fidelity and the detection of leakage.
	We have not performed Surface-17 simulations including~$\ket{3}$ on any qubit, since this increases the simulation cost prohibitively.
	Superleakage is a result of the coherent exchange between~\(\ket{03}\) and~\(\ket{12}\), thus individual events are accompanied by a bit flip on a neighboring qubit.
	The frequency of these events is proportional to the superleakage probability~\(\superleak\).
	Superleakage can result in an increase in the observed defect probabilities, increasing the logical error rate of the code, especially without modifications of the decoder to take this into account~\cite{Kelly15}.
	However, we do not expect superleakage to significantly affect the detection of leakage.
	This is because in the case of a leaked data qubit, the anti-commutation of the neighboring stabilizers still holds, leading to a defect probability of~0.5 regardless of the qubit being in~\(\ket{2}\) or~\(\ket{3}\) (under the assumption that single-qubit gates act trivially on the leakage subspace).
	In the case of a leaked ancilla qubit, the propagated bit flips due to superleakage can be considered as a signature of leakage, in addition to the phase errors due to the leakage conditional phases.

	\section{Leakage steady state in the surface code}
	\label{sec:leakage_steady_state}
	
	Given leakage and seepage probabilities per QEC~cycle, it is expected that each qubit in the surface code equilibrates to a steady-state leakage population after many QEC~cycles.
	Here we do not consider leakage mobility, which is generally small (see~\cref{sec:dynamics_leakage_subspace}), allowing to consider a model for a single qubit.
	We construct a Markovian model to estimate the steady-state populations~$\probcomp$ (resp.~$\probleak$) in the computational subspace~$\compsub$ (leakage subspace~$\leaksub$).
	
	We define~$\Gamma_{i\rightarrow j}$ as the population-transfer probabilities per QEC~cycle.
	The populations are subject to the constraint $\probcomp+\probleak=1$.
	The rate of change of these populations is given by the exchanges from and to each subspace:
	\begin{align}
	\dot{p}^\compsub &= -\probcomp\Gamma_{\compsub\rightarrow \leaksub} + \probleak\Gamma_{\leaksub\rightarrow \compsub}, \nonumber\\ 
	\dot{p}^\leaksub &= \probcomp\Gamma_{\compsub\rightarrow \leaksub}  -\probleak\Gamma_{\leaksub\rightarrow \compsub}. \label{eq:steady_state_differential_equs_2state}
	\end{align}
	The steady-state condition is~$\dot{p}^i=0$ for~$i=\compsub,\leaksub$, resulting in the steady-state populations~$p^i_{ss}$:
	\begin{align}
	\probcomp[ss] &= \frac{\Gamma_{\leaksub\rightarrow \compsub}}{\Gamma_{\compsub\rightarrow \leaksub} + \Gamma_{\leaksub\rightarrow \compsub}}, \nonumber \\
	\probleak[ss] &= \frac{\Gamma_{\compsub\rightarrow \leaksub}}{\Gamma_{\compsub\rightarrow \leaksub} + \Gamma_{\leaksub\rightarrow \compsub}}.
	\end{align}
	
	Considering the $\CZ$~error model in~\cref{sub:leakage_error_model}, for a qubit it approximately holds that
	\begin{align}
	\Gamma_{\compsub\rightarrow \leaksub}&\approx N_\flux\leakrate, \label{eq:gamma12_2state} \\
	\Gamma_{\leaksub\rightarrow \compsub}&\approx N_\flux\seeprate + (1-e^{-\frac{\tcycle}{\Tone/2}}), \label{eq:gamma21_2state} 
	\end{align}
	where $N_\flux$~is in how many $\CZ$~gates the qubit is fluxed during a QEC~cycle, $\tcycle$~is the duration of a \QEC~cycle and~$\leakrate$~(resp.~$\seeprate$) is the average leakage (seepage) probability between~$\compsub$ and~$\leaksub$~\cite{Wood18}.
	The use of the average leakage and seepage probabilities per gate is justified for the surface code because, in the case of data-qubit leakage, ancilla qubits are put in an equal superposition during the parity checks, while, in the case of ancilla-qubit leakage, data qubits are in simultaneous entangled eigenstates of the code stabilizers.
	The seepage probability~[\cref{eq:gamma21_2state}] has one contribution from the unitary $\CZ$-gate interaction and one from relaxation during the entire QEC~cycle.
	Regarding the gate contribution, one has~$\seeprate=2\leakrate$ due to the dimensionality ratio between~$\compsub$ and~$\leaksub$ for a qubit-qutrit pair~\cite{Wood18}.
	
\begin{figure}
		\centering
		\includegraphics[width=\columnwidth]{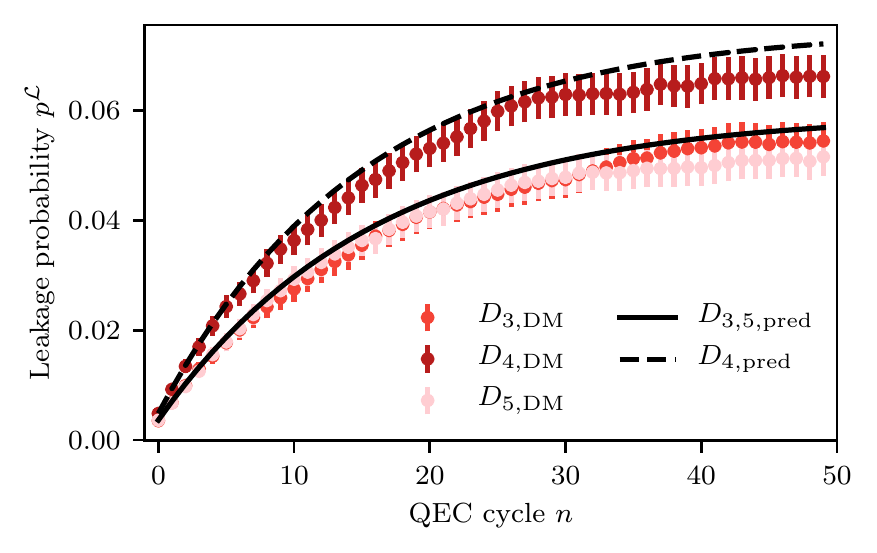}
		\caption{\label{fig:steady_state}
			Evolution of the average leakage population~$\probleak$ towards the steady state over~50 QEC~cycles for the high-frequency data qubits in Surface-17.
			The leakage populations extracted from the density-matrix simulation~(dots) agree well with the predicted one~(black lines).
			The extracted populations are averaged over \(4\times10^{4}\)~runs.
			Error bars correspond to~\(95\%\) confidence intervals estimated by bootstrapping.
		}
	\end{figure}

	The expected steady-state populations in the simulations can be now computed.
	We focus on high-frequency data qubits since the low-frequency ones cannot leak without leakage mobility.
	We have $N_{\CZ}=N_\flux=4$ (for~$\dataq{4}$) or~3 (for~$\dataq{3},\dataq{5}$), $L_1=0.125\%$, $\tcycle=800~\ns$ and~$\Tone=30~\us$.
	The result is~$\probleak[ss]\brkt{\dataq{4}} = 7.5\%$ and~$\probleak[ss]\brkt{\dataq{3}} = \probleak[ss]\brkt{\dataq{5}} = 5.9\%$.
	Furthermore, \cref{eq:steady_state_differential_equs_2state}~can be solved to find that the time evolution of~$\probleak$ towards the steady state is
	\begin{equation}
	\probleak(n) = \frac{\Gamma_{\compsub\rightarrow \leaksub}}{\Gamma_{\compsub\rightarrow \leaksub}+\Gamma_{\leaksub\rightarrow \compsub}} (1-e^{-(\Gamma_{\compsub\rightarrow \leaksub}+\Gamma_{\leaksub\rightarrow \compsub})n}),
	\end{equation}
	where~$n$ is the QEC~cycle number, shown in~\cref{fig:steady_state} for the three high-frequency data qubits.
	We find a good agreement (within error bars) between these predictions and the average leakage population extracted from the density matrix (see~\cref{fig:steady_state}).
	
	We now extend the model to the~$\ket{3}$ state, despite the fact that we have not included it in simulation due to computational constraints.
	To do this, we divide the leakage subspace~$\leaksub$ into the sub-parts~$\leaksub_2$ and~$\leaksub_3$ corresponding to leakage in~$\ket{2}$ and~$\ket{3}$, respectively.
	The rate equations~[\cref{eq:steady_state_differential_equs_2state}] are extended to
	\begin{align}
	\dot{p}^\compsub &= -\probcomp\Gamma_{\compsub \rightarrow \leaksub_2} + p^{\leaksub_2}\Gamma_{\leaksub_2 \rightarrow \compsub}, \nonumber\\ 
	\dot{p}^{\leaksub_2} &= \probcomp\Gamma_{\compsub \rightarrow \leaksub_2} - p^{\leaksub_2}(\Gamma_{\leaksub_2 \rightarrow \compsub} + \Gamma_{\leaksub_2 \rightarrow \leaksub_3}) +  p^{\leaksub_3}\Gamma_{\leaksub_3 \rightarrow \leaksub_2}, \nonumber \\ 
	\dot{p}^{\leaksub_3} &= p^{\leaksub_2}\Gamma_{\leaksub_2 \rightarrow \leaksub_3} -p^{\leaksub_3}\Gamma_{\leaksub_3 \rightarrow \leaksub_2}. \label{eq:steady_state_differential_equs}
	\end{align}
	The steady-state populations $\{p^i_{ss}\}$ then become:
	\begin{align}
	\probcomp[ss] &= \frac{\Gamma_{\leaksub_2 \rightarrow \compsub}\Gamma_{\leaksub_3 \rightarrow \leaksub_2}}{\Gamma_{\compsub \rightarrow \leaksub_2}\Gamma_{\leaksub_3 \rightarrow \leaksub_2} + \Gamma_{\leaksub_2 \rightarrow \compsub}\Gamma_{\leaksub_3 \rightarrow \leaksub_2} + \Gamma_{\compsub \rightarrow \leaksub_2}\Gamma_{\leaksub_2 \rightarrow \leaksub_3}}, \nonumber \\
	p^{\leaksub_2}_{ss} &= \frac{\Gamma_{\compsub \rightarrow \leaksub_2}\Gamma_{\leaksub_3 \rightarrow \leaksub_2}}{\Gamma_{\compsub \rightarrow \leaksub_2}\Gamma_{\leaksub_3 \rightarrow \leaksub_2} + \Gamma_{\leaksub_2 \rightarrow \compsub}\Gamma_{\leaksub_3 \rightarrow \leaksub_2} + \Gamma_{\compsub \rightarrow \leaksub_2}\Gamma_{\leaksub_2 \rightarrow \leaksub_3}}, \nonumber \\
	p^{\leaksub_3}_{ss} &= \frac{\Gamma_{\compsub \rightarrow \leaksub_2}\Gamma_{\leaksub_2 \rightarrow \leaksub_3}}{\Gamma_{\compsub \rightarrow \leaksub_2}\Gamma_{\leaksub_3 \rightarrow \leaksub_2} + \Gamma_{\leaksub_2 \rightarrow \compsub}\Gamma_{\leaksub_3 \rightarrow \leaksub_2} + \Gamma_{\compsub \rightarrow \leaksub_2}\Gamma_{\leaksub_2 \rightarrow \leaksub_3}}.
	\end{align}
	In addition to~\cref{eq:gamma12_2state,eq:gamma21_2state}, in this model we have
	\begin{align}
	\Gamma_{\leaksub_2 \rightarrow \leaksub_3}&\approx N_\flux \superleak/2, \label{eq:gamma23} \\
	\Gamma_{\leaksub_3 \rightarrow \leaksub_2}&\approx N_\flux  \superleak/2 + (1-e^{-\frac{\tcycle}{\Tone/3}}). \label{eq:gamma32}
	\end{align}
	The factor of~$1/2$ in~\cref{eq:gamma23} comes from the fact that superleakage from~$\leaksub_2$ to~$\leaksub_3$ is possible only when the qubit pair performing the~$\CZ$ is in~$\ket{12}$ and not in~$\ket{02}$.
	For~$\superleak=10\%$, for example, the expected steady-state populations are $p^{\leaksub_2}_{ss}\brkt{\dataq{4}} = 7.1\%$, $p^{\leaksub_3}_{ss}\brkt{\dataq{4}} = 5.1\%$ and $p^{\leaksub_2}_{ss}\brkt{\dataq{3}} = p^{\leaksub_2}_{ss}\brkt{\dataq{5}} = 5.7\%$, $p^{\leaksub_3}_{ss}\brkt{\dataq{3}} = p^{\leaksub_3}_{ss}\brkt{\dataq{5}} = 3.8\%$.
	While~$p^{\leaksub_2}_{ss}$ is almost unchanged with respect to the case without superleakage, $p^{\leaksub_3}_{ss}$~has a comparable magnitude to~$p^{\leaksub_2}_{ss}$, suggesting that superleakage needs to be taken into account in optimizing the surface-code performance over many QEC~cycles.

\end{document}